\documentclass[english,onecolumn,conference]{IEEEtran}
\usepackage[T1]{fontenc}
\usepackage[latin9]{inputenc}
\usepackage{amsmath}
\usepackage{amsthm}
\usepackage{amssymb}
\usepackage{esint}

\makeatletter

\providecommand{\tabularnewline}{\\}

\theoremstyle{plain}
\newtheorem{thm}{\protect\theoremname}
\theoremstyle{definition}
\newtheorem{defn}[thm]{\protect\definitionname}
\theoremstyle{plain}
\newtheorem{lem}[thm]{\protect\lemmaname}
\theoremstyle{plain}
\newtheorem{prop}[thm]{\protect\propositionname}
\theoremstyle{remark}
\newtheorem{rem}[thm]{\protect\remarkname}

\usepackage{mathrsfs}
\usepackage{algorithm,algpseudocode}

\usepackage{colortbl}
\definecolor{lightgray}{rgb}{0.9,0.9,0.9}
\definecolor{lightred}{rgb}{1,0.8,0.8}
\definecolor{lightgreen}{rgb}{0.6,1,0.6}
\definecolor{lightyellow}{rgb}{1,1,0.5}
\definecolor{lightgrey}{rgb}{0.8,0.8,0.8}

\allowdisplaybreaks[1]

\makeatother

\usepackage{babel}
\providecommand{\definitionname}{Definition}
\providecommand{\lemmaname}{Lemma}
\providecommand{\propositionname}{Proposition}
\providecommand{\remarkname}{Remark}
\providecommand{\theoremname}{Theorem}

\begin{document}
\title{First-Order Theory of Probabilistic Independence and Single-Letter
Characterizations of Capacity Regions}
\author{Cheuk Ting Li\\
Department of Information Engineering\\
The Chinese University of Hong Kong\\
Email: ctli@ie.cuhk.edu.hk}
\maketitle
\begin{abstract}
We consider the first-order theory of random variables with the probabilistic
independence relation, which concerns statements consisting of random
variables, the probabilistic independence symbol, logical operators,
and existential and universal quantifiers. Although probabilistic
independence is the only non-logical relation included, this theory
is surprisingly expressive, and is able to interpret the true first-order
arithmetic over natural numbers (and hence is undecidable). We also
construct a single-letter characterization of the capacity region
for a general class of multiuser coding settings (including broadcast
channel, interference channel and relay channel), using a first-order
formula. We then introduce the linear entropy hierarchy to classify
single-letter characterizations according to their complexity.
\end{abstract}

\section{Introduction}

In this paper, we study the first-order theory of random variables
with the probabilistic independence relation. We first review some
fragments of the theory studied in the literature. The probabilistic
independence implication problem studied by Geiger, Paz, and Pearl
\cite{geiger1991axioms} and Matúš \cite{matuvs1994stochastic} concerns
the problem of deciding whether a list of probabilistic independence
statements among random variables implies another probabilistic independence
statement, e.g. deciding whether $X\perp\!\!\!\perp Y\,\wedge\,XY\,\perp\!\!\!\perp Z\,\Rightarrow\,X\perp\!\!\!\perp YZ$,
where $\perp\!\!\!\perp$ denotes probabilistic independence, and
juxtaposition $XY$ denotes the joint random variable of $X$ and
$Y$. It was shown in \cite{geiger1991axioms} that probabilistic
independence implication is finitely axiomatizable (the previous example
is one of the axioms), and hence is algorithmically decidable.

The conditional independence implication problem \cite{dawid1979conditional,spohn1980stochastic,mouchart1984note,pearl1987graphoids}
generalizes the probabilistic independence implication problem by
considering probabilistic conditional independence. Pearl and Paz
\cite{pearl1987graphoids} introduced the semi-graphoid axioms, which
was proved to be incomplete by Studen\'y \cite{studeny1989multiinformation}.
As shown in \cite{studeny1990conditional}, no finite axiomization
of probabilistic conditional independence is possible. Nevertheless,
the semi-graphoid axioms are complete for saturated conditional independence
statements \cite{malvestuto1992unique,geiger1993logical}.

It is unknown whether the conditional independence implication problem
is decidable \cite{khamis2020decision}, though some variants of this
problem have been proved to be decidable or undecidable. If the cardinalities
of all random variables are bounded, then it was shown by Niepert
\cite{niepert2012logical} that the problem is decidable (also see
\cite{hannula2019facets}). However, if only the cardinalities of
some random variables are bounded, then it was proved by Li \cite{li2021undecidability}
that the problem is undecidable. Khamis, Kolaitis, Ngo and Suciu \cite{khamis2020decision}
showed that the general conditional independence implication problem
is at most in $\Pi_{1}^{0}$ in the arithmetical hierarchy.

Linear information inequalities \cite{yeung1997framework} concern
linear inequalities among entropy terms on the random variables. Pippenger
\cite{pippenger1986laws} raised the question whether the axiom $I(X;Y|Z)=H(XZ)+H(YZ)-H(XYZ)-H(Z)\ge0$
is sufficient to characterize every true linear information inequality.
This was answered by Zhang and Yeung \cite{zhang1997non,zhang1998characterization}
in the negative, who showed the existence of non-Shannon-type inequalities
not implied by the axiom. More non-Shannon-type inequalities were
given in \cite{makarychev2002new,dougherty2006six,matus2007infinitely,xu2008projection,dougherty2011non}.
Linear information inequalities are closely related to the problem
of finding the capacity region in network coding \cite{yeung2008information,chan2008dualities,yan2012implicit}.
General logical combinations of linear inequalities (using $\wedge$,
$\vee$, $\lnot$) are investigated in \cite{khamis2020decision}.
It is unknown whether the verification of conditional linear information
inequalities is decidable \cite{dougherty2009undecidable,gomez2014network,khamis2020decision},
though if the problem is extended to allow affine inequalities, then
it was shown in \cite{li2021undecidability} that the problem is undecidable.

While all aforementioned problems are not existential (they are purely
universal statements in the form $\forall X^{n}.P(X^{n})$, where
$P$ is a predicate, and all random variables $X^{n}=(X_{1},\ldots,X_{n})$
are universally quantified, i.e., they are in the $\forall^{*}$-fragment
of the first-order theory of random variables), existential results
on random variables (concerning predicates on $X^{n}$ in the form
$\exists U^{m}.P(X^{n},U^{m})$) are widely used in information theory.
For example, in network information theory \cite{elgamal2011network},
capacity regions are often expressed as statements concerning the
existence of some auxiliary random variables. Some examples of useful
existential formulae include the double Markov property \cite{csiszar2011information}
($X\perp\!\!\!\perp Z|Y$ denotes conditional independence)
\begin{align*}
 & X\perp\!\!\!\perp Z|Y\,\wedge\,Y\perp\!\!\!\perp Z|X\\
 & \Rightarrow\exists U.(U\perp\!\!\!\perp U|X\,\wedge\,U\perp\!\!\!\perp U|Y\,\wedge\,XY\perp\!\!\!\perp Z|U),
\end{align*}
the copy lemma \cite{zhang1998characterization,dougherty2011non},
and the functional representation lemma \cite{elgamal2011network}
(also see \cite{sfrl_trans}). The existential theory of random variables
with information inequalities has been studied systematically in \cite{li2021automated}.

There are examples of rate regions and bounds in network information
theory expressed in a nested ``for all, there exists'' form (i.e.,
a predicate on $X^{n}$ in the form $\forall Y^{k}.P(X^{n},Y^{k})\to\exists U^{m}.Q(X^{n},Y^{k},U^{m})$),
e.g. the outer bound for multiterminal source coding in \cite{wagner2008improved},
the upper bound on key agreement in \cite[Cor. 2]{gohari2010information},
the outer bound for the source broadcast problem in \cite[Thm 1]{yu2018distortion},
and the auxiliary receivers in \cite{gohari2020outer}. In these examples,
there are both existentially and universally quantified auxiliary
random variables. They can be considered as formulae in (the $\forall^{*}\exists^{*}$-fragment
of) the first-order theory of random variables, with a larger depth
of quantifier alternation compared to existential formulae.

Although the aforementioned special cases have been studied extensively,
to the best of the author's knowledge, there has not been a systematic
treatment on the general first-order theory of random variables with
arbitrary depth of quantifier alternation (refer to the related work
section). In this paper, we investigate the \emph{first-order theory
of random variables with probabilistic independence} (\emph{FOTPI}),
which concerns formulae consisting of variables (which represents
random variables), the probabilistic independence symbol $\perp\!\!\!\perp$,
logical operators ($\wedge$, $\vee$, $\lnot$) and existential and
universal quantifiers ($\exists,\forall$).\footnote{We remark that this paper does not take an axiomatic approach. Theorems
are not derived from a finite set of axioms of the first-order theory
of probabilistic independence, but rather from the underlying model
of random variables in a probability space (i.e., the axioms of the
theory are taken to be the set of true first-order sentences about
random variables). Since FOTPI can interpret the true first-order
arithmetic, by Gödel's first incompleteness theorem, FOTPI is not
recursively axiomatizable.} Even though probabilistic independence is the only non-logical relation
included, we can use it to define other concepts in probability such
as conditional independence, functional dependency, uniformity, cardinality
and entropy. Therefore, most of the aforementioned problems can be
considered as fragments of FOTPI. Since cardinality bounds can be
defined using first-order formulae, as a corollary of \cite{li2021undecidability},
FOTPI is undecidable. Also, we show that FOTPI can interpret the true
first-order arithmetic over natural numbers. 

Furthermore, we prove that for any setting within a general class
of multiuser coding settings (including broadcast channel \cite{cover1972broadcast},
interference channel \cite{ahlswede1974capacity}, relay channel \cite{van1971three},
and finite-state Markov channel \cite{goldsmith1996capacity} \footnote{This is perhaps surprising, considering that the capacity of the finite-state
Markov channel is uncomputable \cite{elkouss2018memory,boche2020shannon}.
This is used in \cite{agarwal2018non} to show that the capacity of
the finite-state Markov channel does not admit a single-letter characterization
in a certain form.}), the capacity region has a single-letter characterization that can
be stated as a formula in FOTPI (here ``single-letter'' means the
number of random variables in the formula is fixed). Whether this
can be regarded as a solution to the open problem of finding single-letter
characterizations of the capacity regions of broadcast channel and
interference channel depends on the definition of ``single-letter
characterization''. While there is no generally accepted definition
of what constitutes a single-letter characterization \cite{korner1987concept},
it can be argued that if ``for all, there exists'' statements (e.g.
\cite{wagner2008improved,gohari2010information,yu2018distortion,gohari2020outer})
are considered single-letter, then there is no reason to exclude statements
with a larger depth of quantifier alternation. 

The single-letter characterization given in this paper is very complex
(its depth of quantifier alternation is $17$), which is against the
purpose of single-letter characterizations. Perhaps, instead of asking
for any single-letter characterization of the capacity region of the
broadcast/interference/relay channel, a more precise question is to
find the simplest single-letter characterization, which will hopefully
provide more insight to the optimal coding scheme\footnote{Note that even purely existential predicates (which include most existing
single-letter formulae for capacity regions) can be undecidable without
cardinality bounds. See \cite{li2021undecidability} and Proposition
\ref{prop:h_frag}. Therefore, simplicity of a formula does not necessarily
imply ease of computation.}. We propose a classification of first-order formulae, called the
\emph{linear entropy hierarchy}, according to their depth of quantifier
alternation. We can then rigorously state the open problems of finding
single-letter characterizations of capacity regions of the aforementioned
channels in the lowest possible level in the linear entropy hierarchy.

This paper is organized as follows. In Section \ref{sec:relations},
we define some relations over random variables in FOTPI. In Section
\ref{sec:encint}, we show that FOTPI can interpret the first-order
arithmetic. In Section \ref{sec:definable}, we investigate definability
in FOTPI. In Section \ref{sec:event}, we study a representation of
events in FOTPI. In Section \ref{sec:seq}, we study a representation
of random sequences in FOTPI. In Section \ref{sec:singleletter},
we present the main result, which is a single-letter characterization
of the capacity region for a general class of multiuser coding settings.
In Section \ref{sec:defslc}, we propose the linear entropy hierarchy
as a classification of first-order formulae according to their complexity.
In Section \ref{sec:continuous}, we study the extension to continuous
random variables.

\medskip{}

\subsection{Related Work}

A computer program called PSITIP is described in \cite{li2021automated},
which is capable of expressing and verifying some first-order statements
on random variables, though the paper \cite{li2021automated} is focused
only on existential and implication problems. 

Regarding inner bounds of capacity regions of multiuser coding settings,
several general inner bounds were studied in \cite{rini2013general,minero2015unified,lee2015unified,lee2018unified}.
These bounds can be regarded as purely existential formulae in FOTPI.
The author is unaware of any general result on outer bounds of multiuser
coding settings, though Gallager's strategy \cite{gallager1974capacity},
a standard method for proving outer bounds, can be applied automatically
by the PSITIP software \cite{li2021automated} to discover outer bounds
for general multiuser settings. The resultant bounds are also purely
existential.

Khamis, Kolaitis, Ngo and Suciu \cite{khamis2020decision} studied
several classes of statements about conditional independence and linear
inequalities on entropy (which are all purely universal statements),
and gave upper-bounds on their hardness in the arithmetical hierarchy,
which is, loosely speaking, characterized by the depth of quantifier
alternation in a first-order formula on natural numbers. In other
words, \cite{khamis2020decision} attempted to express purely universal
first-order formulae on random variables as first-order formulae on
natural numbers with the lowest depth of quantifier alternation (general
first-order formulae on random variables are not studied in \cite{khamis2020decision}).
In comparison, this paper tries to express capacity regions as first-order
formulae on random variables with the lowest depth of quantifier alternation. 

Another undecidability result in information theory is the capacity
of the finite-state Markov channel \cite{elkouss2018memory,boche2020shannon}.
Using this fact, Agarwal \cite{agarwal2018non} showed that the capacity
of the finite-state Markov channel does not admit a single-letter
characterization expressible as a conjunction of linear inequalities
in the form $R\le\sum_{i}\alpha_{i}I(U_{A_{i}};U_{B_{i}}|U_{C_{i}})$
($U_{A_{i}}=\{U_{a}\}_{a\in A_{i}}$, $A_{i},B_{i},C_{i}\subseteq[n]$)
and polynomial constraints on the joint probability mass function,
where the alphabets of all auxiliary random variables are fixed. This
definition of single-letter characterization shares some similarities
to the first existential level of the linear entropy hierarchy in
this paper (refer to Section \ref{sec:defslc} and Remark \ref{rem:prev}
for the similarities and differences).

We also remark that the graphoid and separoid axioms are expressed
using first-order languages in the work by Córdoba-Sánchez, Bielza,
and Larranaga \cite{cordoba2016graphoids}, though \cite{cordoba2016graphoids}
has not studied the expressive power of a language with only probabilistic
independence.

The notion of normalized entropy vectors was introduced in \cite{hassibi2007normalized},
which used it to give a characterization of the capacity region of
a communication network, which is single-letter in a certain sense.
However, it appears that representing a discrete memoryless multiuser
channel in the setting in \cite{hassibi2007normalized} is not entirely
straightforward \footnote{It appears that the channel constraint in \cite{hassibi2007normalized}
is for one-shot communication. While one may apply \cite{hassibi2007normalized}
on the infinite product channel, and use the cardinality of the output
to normalize the rate against the number of channel uses, such approach
would not be considered single-letter in the usual sense.}. 

Note that FOTPI in this paper is not the same as the first-order probabilistic
logic \cite{nilsson1986probabilistic,koller1996irrelevance,cozman2008probabilistic},
which is used in knowledge representation. We are interested in the
information contained in the random variables instead of their values
(i.e., random variables are considered equivalent under relabelling),
whereas first-order probabilistic logic concerns the knowledge on
the values of the random variables.

\medskip{}

\subsection*{Notations}

We write $\mathbb{N}_{+}:=\{1,2,\ldots\}$, $\mathbb{N}_{0}:=\{0,1,2,\ldots\}$,
$[a..b]:=[a,b]\cap\mathbb{Z}$, $[n]:=[1..n]$. The uniform distribution
over the set $S$ is denoted as $\mathrm{Unif}(S)$. The Bernoulli
distribution is denoted as $\mathrm{Bern}(a)$ (which is $1$ with
probability $a$, $0$ with probability $1-a$).

\medskip{}

\section{Relations over Random Variables\label{sec:relations}}

For simplicity, all random variables are assumed to be discrete unless
otherwise stated (the case for general random variables is discussed
in Section \eqref{sec:continuous}). Since we only consider discrete
random variables, we may assume that they are all defined in the standard
probability space $([0,1],\mathcal{F},P)$ (where $P$ is the Lebesgue
measure, and $\mathcal{F}$ is the $\sigma$-algebra of Lebesgue measurable
subsets of $[0,1]$). Therefore, in this paper, the set of all random
variables is taken to be the set of measurable functions from $[0,1]$
to $\mathbb{N}_{+}$. Let this set be $\mathcal{M}$. Since the labelling
of the random variable does not matter in the probabilistic independence
relation, we may also regard a random variable as a finite or countably-generated
$\sigma$-subalgebra of $\mathcal{F}$, though we will use random
variables instead of $\sigma$-subalgebras in this paper for notational
simplicity.

The FOTPI is the first-order theory of $(\mathcal{M},\perp\!\!\!\perp)$,
where $\perp\!\!\!\perp$ stands for probabilistic independence between
two random variables. This theory consists of all first-order sentences
$\psi$ where $\mathcal{M}\models\psi$ (i.e., it is the complete
theory $\mathrm{Th}(\mathcal{M},\perp\!\!\!\perp)$ of the system
of random variables $\mathcal{M}$).

We write $X\stackrel{\iota}{\le}Y$ for the condition that $X$ is
(almost surely) a function of $Y$ (i.e., there exists a function
$f$ such that $X=f(Y)$ almost surely). This can be expressed using
independence as
\begin{equation}
X\stackrel{\iota}{\le}Y:=\,\forall U.\,\big(U\perp\!\!\!\perp Y\,\to\,U\perp\!\!\!\perp X\big).\label{eq:lei}
\end{equation}
It is clear that $X\stackrel{\iota}{\le}Y$ implies $\forall U:\,(U\perp\!\!\!\perp Y\,\to\,U\perp\!\!\!\perp X)$.
For the other direction, if $X\stackrel{\iota}{\le}Y$ does not hold,
then there exists $x_{0},y_{0}$ such that $0<p_{X|Y}(x_{0}|y_{0})<1$.
Let $U\in\{0,1\}$ be a random variable such that 
\[
p_{U|X,Y}(1|x,y)=\begin{cases}
1 & \mathrm{if}\;y=y_{0},\,x=x_{0}\\
0 & \mathrm{if}\;y=y_{0},\,x\neq x_{0}\\
p_{X|Y}(x_{0}|y_{0}) & \mathrm{if}\;y\neq y_{0}.
\end{cases}
\]
It is clear that $U\perp\!\!\!\perp Y$ and $U\not\perp\!\!\!\perp X$.

We write $X\stackrel{\iota}{=}Y$ for the condition that $X$ is informationally
equivalent to $Y$ (i.e., there exists an injective function $f$
such that $X=f(Y)$ almost surely). This can be expressed as
\[
X\stackrel{\iota}{=}Y:=\,X\stackrel{\iota}{\le}Y\,\wedge\,Y\stackrel{\iota}{\le}X.
\]

We use juxtaposition $XY$ to denote the joint random variable of
$X$ and $Y$ (while we assume random variables take values over natural
numbers, we may apply any bijection to pairs of values of $X,Y$ to
the natural numbers, since any two bijections are equivalent under
$\stackrel{\iota}{=}$). Joint random variable can be characterized
by the lattice join operation \cite{li2011connection}:
\begin{align}
 & Z\stackrel{\iota}{=}XY\,\nonumber \\
 & \Leftrightarrow\,X\stackrel{\iota}{\le}Z\,\wedge\,Y\stackrel{\iota}{\le}Z\nonumber \\
 & \;\;\;\;\wedge\,\forall U.\,\big((X\stackrel{\iota}{\le}U\,\wedge\,Y\stackrel{\iota}{\le}U)\,\to\,Z\stackrel{\iota}{\le}U\big).\label{eq:joint}
\end{align}
Therefore, it is not necessary to include joint random variable in
the language of FOTPI, though we would still use the notation $XY$
for the $Z$ satisfying the above formula for notational simplicity.

Mutual independence among random variables $X_{1},\ldots,X_{n}$ can
be expressed as
\begin{align*}
 & X_{1}\perp\!\!\!\perp\cdots\perp\!\!\!\perp X_{n}\\
 & \Leftrightarrow\bigwedge_{i=2}^{n}\big(X_{i}\perp\!\!\!\perp(X_{1}\cdots X_{i-1})\big).
\end{align*}

Write $X\perp\!\!\!\perp Y|Z$ for the condition that $X,Y$ are conditionally
independent given $Z$. Using the functional representation lemma
\cite{elgamal2011network}, we can express conditional independence
as
\begin{align}
 & X\perp\!\!\!\perp Y|Z\nonumber \\
 & \Leftrightarrow\;\exists U.\,U\perp\!\!\!\perp XZ\,\wedge\,Y\stackrel{\iota}{\le}ZU.\label{eq:ci}
\end{align}
We also define
\begin{align*}
X\stackrel{\iota}{=}Y & :=\,X\stackrel{\iota}{\le}Y\,\wedge\,Y\stackrel{\iota}{\le}X,\\
X\stackrel{\iota}{\neq}Y & :=\,\lnot(X\stackrel{\iota}{=}Y),\\
X\stackrel{\iota}{<}Y & :=\,X\stackrel{\iota}{\le}Y\,\wedge\,\lnot(Y\stackrel{\iota}{\le}X).
\end{align*}

\medskip{}

\section{Representation of Integers\label{sec:encint}}

In this section, we describe a representation of integers as random
variables. We show that the first-order theory of arithmetic over
nonnegative integers is interpretable in the first-order theory of
probabilistic independence. 

\subsection{Uniformity}

To test whether $X$ is uniformly distributed over its support, we
use the result in \cite{zhang1997non} that if $X,Y,Z$ are discrete
random variables such that any one of them is a function of the other
two, and they are pairwise independent, then they are all uniformly
distributed over their supports, which have the same size. Using the
notations in \cite{li2021undecidability}, the condition that $X$
is uniformly distributed over its support can be expressed as
\[
\mathrm{unif}(X):=\exists Y,Z.\,\mathrm{triple}(X,Y,Z),
\]
where
\begin{align*}
\mathrm{triple}(X,Y,Z):= & X\stackrel{\iota}{\le}YZ\,\wedge\,Y\stackrel{\iota}{\le}XZ\,\wedge\,Z\stackrel{\iota}{\le}XY\\
 & \wedge\,X\perp\!\!\!\perp Y\,\wedge\,X\perp\!\!\!\perp Z\,\wedge\,Y\perp\!\!\!\perp Z.
\end{align*}

\subsection{Cardinality}

We write $\mathcal{X}$ for the support of $X$, and $|\mathcal{X}|$
for the cardinality of $X$. To test whether $X$ is (at most) a binary
random variable (i.e., $|\mathcal{X}|\le2$), note that any random
variable with strictly less information than $X$ must be degenerate,
and hence the condition that $X$ is (at most) a binary random variable
can be expressed as
\[
\mathrm{card}_{\le2}(X):=\forall U\big(U\stackrel{\iota}{<}X\,\to\,U\stackrel{\iota}{=}\emptyset\big).
\]
By $U\stackrel{\iota}{=}\emptyset$, we mean that $U$ is informationally
equivalent to the constant random variable. This can be expressed
without introducing a new constant $\emptyset$ by $U\stackrel{\iota}{=}\emptyset\,\Leftrightarrow\,U\perp\!\!\!\perp U$.

If $X$ has cardinality at most $n$, then any random variable with
strictly less information than $X$ has cardinality at most $n-1$.
Therefore, the condition that $|\mathcal{X}|\le n$ ($n\ge2$) can
be defined recursively as
\begin{align}
\mathrm{card}_{\le n}(X) & :=\forall U\big(U\stackrel{\iota}{<}X\,\to\,\mathrm{card}_{\le n-1}(U)\big),\label{eq:cardlen}\\
\mathrm{card}_{\le1}(X) & :=(X\stackrel{\iota}{=}\emptyset).\nonumber 
\end{align}
We can then define 
\begin{equation}
\mathrm{card}_{=n}(X):=\mathrm{card}_{\le n}(X)\,\wedge\,\lnot\mathrm{card}_{\le n-1}(X),\label{eq:cardeq}
\end{equation}
\[
\mathrm{card}_{\ge n}(X):=\lnot\mathrm{card}_{\le n-1}(X).
\]

\medskip{}

\subsection{Relations over Integers}

Given the tests for uniformity and cardinality, a natural way to represent
a positive integer $k$ as a random variable is to represent it as
a uniformly distributed random variable with cardinality $k$. In
this section, we express several relations over positive integers
using first-order formulae. Note that some of these formulae have
appeared in \cite{li2021undecidability}.
\begin{itemize}
\item (Equality) The formula for checking $|\mathcal{X}|=|\mathcal{Y}|$
for uniform $X,Y$ is given by \cite{li2021undecidability}:
\begin{align*}
\mathrm{ueq}(X,Y) & :=\exists U_{1},U_{2},U_{3}.\,\\
 & \;\;\;\;\mathrm{triple}(X,U_{1},U_{2})\,\wedge\,\mathrm{triple}(Y,U_{1},U_{3}).
\end{align*}
We also define
\[
\mathrm{ueq}_{n}(X):=\mathrm{unif}(X)\,\wedge\,\mathrm{card}_{=n}(X)
\]
to check for equality against constants.
\item (Multiplication) The formula for checking $|\mathcal{Z}|=|\mathcal{X}||\mathcal{Y}|$
for uniform $X,Y$ is given by \cite{li2021undecidability}:
\begin{align*}
\mathrm{uprod}(X,Y,Z) & :=\exists\tilde{X},\tilde{Y}.\,\big(\mathrm{ueq}(X,\tilde{X})\,\wedge\,\mathrm{ueq}(Y,\tilde{Y})\\
 & \;\;\;\wedge\,\tilde{X}\perp\!\!\!\perp\tilde{Y}\,\wedge\,\tilde{X}\tilde{Y}\stackrel{\iota}{=}Z\big).
\end{align*}
\item (Comparison) The formula for checking $|\mathcal{X}|\le|\mathcal{Y}|$
for uniform $X,Y$ is given by \cite{li2021undecidability} (with
slight modification):
\begin{align*}
 & \mathrm{ule}(X,Y)\\
 & :=\exists G,\tilde{Y}.\,\big(\mathrm{uprod}(X,Y,G)\,\wedge\,\mathrm{ueq}(Y,\tilde{Y})\,\wedge\,G\stackrel{\iota}{\le}Y\tilde{Y}\big).
\end{align*}
We briefly repeat the reason given in \cite{li2021undecidability}
here. Note that $\mathrm{uprod}(X,Y,G)\,\wedge\,\mathrm{ueq}(Y,\tilde{Y})\,\wedge\,G\stackrel{\iota}{\le}Y\tilde{Y}$
implies $|\mathcal{X}||\mathcal{Y}|=|\mathcal{G}|\le|\mathcal{Y}|^{2}$,
which implies $|\mathcal{X}|\le|\mathcal{Y}|$. For the other direction,
assume $\mathcal{X}=\{0,\ldots,|\mathcal{X}|-1\}$, $\mathcal{Y}=\{0,\ldots,|\mathcal{Y}|-1\}$,
$|\mathcal{X}|\le|\mathcal{Y}|$. Take $G=(Y,\tilde{X})$, where $\tilde{X}\sim\mathrm{Unif}(\mathcal{X})$
is independent of $Y$. Take $\tilde{Y}=\tilde{X}+Y\;\mathrm{mod}\,|\mathcal{Y}|$.
It is clear that $G\stackrel{\iota}{\le}Y\tilde{Y}$. Therefore, the
formula for strict inequality $|\mathcal{X}|<|\mathcal{Y}|$ is
\[
\mathrm{ult}(X,Y):=\lnot\mathrm{ule}(Y,X).
\]
Also define $\mathrm{uge}$ and $\mathrm{ugt}$ similarly. We define
\[
\mathrm{ule}_{n}(X):=\exists U.\,\mathrm{ueq}_{n}(U)\,\wedge\,\mathrm{ule}(X,U),
\]
and similar for $\mathrm{ult}_{n}$, $\mathrm{uge}_{n}$, $\mathrm{ugt}_{n}$.
\item (Divisibility) The condition that $|\mathcal{Y}|$ is divisible by
$|\mathcal{X}|$ for uniform $X,Y$ can be expressed as 
\[
\mathrm{udiv}(X,Y):=\exists U.\,\mathrm{uprod}(X,U,Y).
\]
The condition that $|\mathcal{X}|$ is a prime number is
\[
\mathrm{uprime}(X):=\lnot\exists U,V.\,\big(U\stackrel{\iota}{\neq}\emptyset\,\wedge\,V\stackrel{\iota}{\neq}\emptyset\,\wedge\,\mathrm{uprod}(U,V,X)\big).
\]
\item (Successor) The condition that $|\mathcal{Y}|=|\mathcal{X}|+1$ for
uniform $X,Y$ can be expressed as 
\[
\mathrm{usucc}(X,Y):=\mathrm{ult}(X,Y)\,\wedge\,\forall U.\big(\mathrm{ult}(X,U)\to\mathrm{ule}(Y,U)\big).
\]
\end{itemize}

\medskip{}

\subsection{Interpreting First-order Arithmetic}

In order to interpret the first-order theory of arithmetic, it is
left to define addition. Given a discrete random variable $X$, we
call $Y$ a \emph{single-mass indicator} of $X$ if there exists $x$
such that $\mathbf{P}(X=x)>0$ and $Y\stackrel{\iota}{=}\mathbf{1}_{\{x\}}(X)$
(i.e., $Y$ is the indicator function of $X=x$). This can be characterized
by the first-order formula
\begin{align}
\mathrm{smi}(X,Y) & :=(X\stackrel{\iota}{=}Y\stackrel{\iota}{=}\emptyset)\nonumber \\
 & \;\;\;\vee\big(Y\stackrel{\iota}{\le}X\,\wedge\,\mathrm{card}_{=2}(Y)\,\nonumber \\
 & \;\;\;\;\;\wedge\,\forall U.\big(U\stackrel{\iota}{\le}X\,\wedge\,\mathrm{card}_{=4}(U)\nonumber \\
 & \;\;\;\;\;\;\;\to\lnot\exists V.(\mathrm{card}_{\le2}(V)\,\wedge\,U\stackrel{\iota}{\le}YV)\big)\big).\label{eq:smi}
\end{align}
To check this, note that if $Y\stackrel{\iota}{=}\mathbf{1}_{\{x\}}(X)$
and $U\stackrel{\iota}{\le}X\,\wedge\,\mathrm{card}_{=4}(U)$, then
there are at least $3$ possible values of $U$ given $Y=0$, and
there does not exist $V$ with $\mathrm{card}_{\le2}(V)\,\wedge\,U\stackrel{\iota}{\le}YV$.
For the other direction, assume $Y\in\{0,1\}$ is binary but is not
a single-mass indicator of $X$. We can therefore construct $U\stackrel{\iota}{\le}X$
which takes two possible values given $Y=0$ or $Y=1$, and there
exists $V$ with $\mathrm{card}_{\le2}(V)\,\wedge\,U\stackrel{\iota}{\le}YV$
(which indicates which of the two values $U$ takes).

Consider
\begin{align}
\mathrm{frac}(X,Y,Z,U) & :=\big(\mathrm{ueq}_{2}(U)\,\wedge\,\mathrm{uprod}(X,U,Z)\,\wedge\,\mathrm{uprod}(Y,U,Z)\big)\nonumber \\
 & \;\;\;\vee\exists\tilde{X},\tilde{Y}.\,\big(\mathrm{ueq}(X,\tilde{X})\,\wedge\,\mathrm{ueq}(Y,\tilde{Y})\,\wedge\,\mathrm{unif}(Z)\,\nonumber \\
 & \;\;\;\;\;\;\wedge\,\mathrm{card}_{=2}(U)\,\wedge\,\lnot\mathrm{unif}(U)\nonumber \\
 & \;\;\;\;\;\;\wedge\,U\stackrel{\iota}{\le}Z\,\wedge\,\tilde{X}\perp\!\!\!\perp\tilde{Y}\perp\!\!\!\perp U\,\wedge\,Z\stackrel{\iota}{\le}\tilde{X}\tilde{Y}U\nonumber \\
 & \;\;\;\;\;\;\wedge\,\forall V.\big(\mathrm{smi}(Z,V)\,\to\,\mathrm{smi}(\tilde{X}U,V)\vee\mathrm{smi}(\tilde{Y}U,V)\big)\big).\label{eq:usum}
\end{align}
We will show that $\mathrm{frac}(X,Y,Z,U)$ holds if and only if $X,Y,Z$
are uniform, $|\mathcal{Z}|=|\mathcal{X}|+|\mathcal{Y}|$, and $U\sim\mathrm{Bern}(|\mathcal{X}|/(|\mathcal{X}|+|\mathcal{Y}|))$.
Note that the first case $\mathrm{ueq}_{2}(U)\,\wedge\,\mathrm{uprod}(X,U,Z)\,\wedge\,\mathrm{uprod}(Y,U,Z)$
checks for $|\mathcal{X}|=|\mathcal{Y}|=|\mathcal{Z}|/2$. The second
case checks for $|\mathcal{Z}|=|\mathcal{X}|+|\mathcal{Y}|$, $|\mathcal{X}|\neq|\mathcal{Y}|$.
To check the second case, note that if $|\mathcal{Z}|=|\mathcal{X}|+|\mathcal{Y}|$
(assume $\mathcal{X}=[|\mathcal{X}|]$ and the same for $\mathcal{Y},\mathcal{Z}$),
then we can let $U\sim\mathrm{Bern}(|\mathcal{X}|/(|\mathcal{X}|+|\mathcal{Y}|))$,
$\tilde{X}\sim\mathrm{Unif}[|\mathcal{X}|]$, $\tilde{Y}\sim\mathrm{Unif}[|\mathcal{Y}|]$,
$\tilde{X}\perp\!\!\!\perp\tilde{Y}\perp\!\!\!\perp U$ and $Z=\tilde{X}$
if $U=1$, $Z=\tilde{Y}+|\mathcal{X}|$ if $U=0$. It is straightforward
to check $Z\sim\mathrm{Unif}[|\mathcal{X}|+|\mathcal{Y}|]$, $Z\stackrel{\iota}{\le}\tilde{X}\tilde{Y}U$
and $\forall V.(\mathrm{smi}(Z,V)\,\to\,\mathrm{smi}(\tilde{X}U,V)\vee\mathrm{smi}(\tilde{Y}U,V))$. 

For the other direction, assume $\mathrm{frac}(X,Y,Z,U)$ holds and
$|\mathcal{X}|\neq|\mathcal{Y}|$. Let $U\sim\mathrm{Bern}(\theta)$,
$\theta\in(0,1)\backslash\{1/2\}$. Note that if $\mathrm{smi}(Z,V)$,
then $V\sim\mathrm{Bern}(1/|\mathcal{Z}|)$. Also, $\mathrm{smi}(\tilde{X}U,V)\vee\mathrm{smi}(\tilde{Y}U,V)$
holds only if $V\sim\mathrm{Bern}(\phi)$ where $\phi=\theta/|\mathcal{X}|$,
$\theta/|\mathcal{Y}|$, $(1-\theta)/|\mathcal{X}|$, or $(1-\theta)/|\mathcal{Y}|$.
Since $\theta/|\mathcal{X}|\neq(1-\theta)/|\mathcal{X}|$, we either
have $1/|\mathcal{Z}|=\theta/|\mathcal{X}|=(1-\theta)/|\mathcal{Y}|$,
$1/|\mathcal{Z}|=\theta/|\mathcal{X}|=1-(1-\theta)/|\mathcal{Y}|$
or $1/|\mathcal{Z}|=1-\theta/|\mathcal{X}|=1-(1-\theta)/|\mathcal{Y}|$
(the other cases are similar by symmetry). The first case gives $\theta=|\mathcal{X}|/(|\mathcal{X}|+|\mathcal{Y}|)$,
$|\mathcal{Z}|=|\mathcal{X}|+|\mathcal{Y}|$. For the second case,
it implies $|\mathcal{Z}|\ge2$, and hence $|\mathcal{Y}|=1$. Since
$\theta/|\mathcal{X}|=1-(1-\theta)/|\mathcal{Y}|$, we have $|\mathcal{X}|=1$.
Since $Z\stackrel{\iota}{\le}\tilde{X}\tilde{Y}U$, we have $|\mathcal{Z}|=2$,
and $U\sim\mathrm{Bern}(1/2)$, giving a contradiction. For the third
case, it implies $|\mathcal{Z}|\ge2$, $|\mathcal{X}|=|\mathcal{Y}|=1$.
Since $1-\theta/|\mathcal{X}|=1-(1-\theta)/|\mathcal{Y}|$, we have
$\theta=1/2$, giving a contradiction. Hence, only the first case
is possible, and $|\mathcal{Z}|=|\mathcal{X}|+|\mathcal{Y}|$.

We can therefore define addition as follows. The formula for checking
$|\mathcal{Z}|=|\mathcal{X}|+|\mathcal{Y}|$ for uniform $X,Y,Z$
is given by
\[
\mathrm{usum}(X,Y,Z):=\exists U.\mathrm{frac}(X,Y,Z,U).
\]

We now show that true arithmetic \cite{boolos2002computability} (i.e.,
the theory $\mathrm{Th}(\mathbb{N}_{0},+,\cdot,<)$ containing all
true first-order sentences over nonnegative integers with addition,
multiplication and comparison) is interpretable in the first-order
theory of probabilistic independence. 
\begin{thm}
\label{thm:interpret}True arithmetic is interpretable in the first-order
theory of probabilistic independence. 
\end{thm}
\begin{IEEEproof}
We represent $a\in\mathbb{N}_{+}$ by a uniform random variable with
cardinality $a$. Note that true arithmetic concerns $\mathbb{N}_{0}$
instead of $\mathbb{N}_{+}$, so we need a special representation
for $0$. We represent $0$ by a random variable with distribution
$\mathrm{Bern}(1/3)$ (up to relabeling). This distribution can be
checked by observing that $X\sim\mathrm{Bern}(1/3)$ (up to relabeling)
if and only if
\[
\mathrm{is0}(X):=\exists U.(\mathrm{ueq}_{3}(U)\,\wedge\,\emptyset\stackrel{\iota}{<}X\stackrel{\iota}{<}U).
\]
The following formula checks whether $X$ is the representation of
an integer in $\mathbb{N}_{0}$:
\begin{equation}
\mathrm{isnat}(X):=\mathrm{is0}(X)\,\vee\,\mathrm{unif}(X).\label{eq:isnat}
\end{equation}
It is straightforward to modify the definitions of $\mathrm{usum}$
and $\mathrm{uprod}$ to accommodate this special value of $0$.

\end{IEEEproof}
\medskip{}

As a result, by Tarski's undefinability theorem \cite{tarski1933pojkecie,boolos2002computability},
FOTPI is not arithmetically definable.

\medskip{}

\section{Definable Distributions\label{sec:definable}}

In this section, we investigate the concept of definability in FOTPI.
\begin{defn}
[Definability]We use the following definitions of definability:
\begin{itemize}
\item (Definability of distributions) We call a probability mass function
$p$\emph{ definable} in FOTPI if there exists a first-order formula
$P(X)$ such that $P(X)$ holds if and only if $X$ follows the distribution
$p$ up to relabeling (i.e., there exists an injective function $f$
such that $f(X)\sim p$). We call a set $S\subseteq\mathcal{M}^{n}$
definable in FOTPI if there exists a first-order formula $P(X_{1},\ldots,X_{n})$
which holds if and only if $(X_{1},\ldots,X_{n})\in S$ (note that
a relation is a special case of a set for $n=2$).\medskip{}
\item (Bernoulli definability over reals) We call a real number $\theta\in[0,1/2]$
\emph{Bernoulli-definable} in FOTPI if the probability mass function
of the Bernoulli distribution $\mathrm{Bern}(\theta)$ is definable
in FOTPI. We call a set $S\subseteq[0,1/2]^{n}$ Bernoulli-definable
in FOTPI if there exists a first-order formula $P(X_{1},\ldots,X_{n})$
which holds if and only if $X_{i}\sim\mathrm{Bern}(\theta_{i})$ (up
to relabelling), $\theta_{i}\in[0,1/2]$, and $(\theta_{1},\ldots,\theta_{n})\in S$.
For $S\subseteq[0,1/2]^{n}$, we call a function $f:S\to[0,1/2]$
Bernoulli-definable in FOTPI if its graph $\{(\theta_{1},\ldots,\theta_{n},f(\theta_{1},\ldots,\theta_{n})):\,(\theta_{1},\ldots,\theta_{n})\in S\}\in[0,1/2]^{n+1}$
is Bernoulli-definable in FOTPI.\medskip{}
\item (Uniform definability over natural numbers) We call a set $S\subseteq\mathbb{N}_{+}^{n}$
\emph{uniform-definable} in FOTPI if there exists a first-order formula
$P(X_{1},\ldots,X_{n})$ which holds if and only if $X_{i}\sim\mathrm{Unif}[k_{i}]$
(up to relabelling) and $(k_{1},\ldots,k_{n})\in S$. For $S\subseteq\mathbb{N}_{+}^{n}$,
we call a function $f:S\to\mathbb{N}_{+}$ uniform-definable in FOTPI
if its graph $\{(\theta_{1},\ldots,\theta_{n},f(\theta_{1},\ldots,\theta_{n})):\,(\theta_{1},\ldots,\theta_{n})\in S\}\in\mathbb{N}_{+}^{n+1}$
is uniform-definable in FOTPI.
\end{itemize}
\end{defn}
\medskip{}

Since 
\begin{equation}
\mathrm{qeq}(X,Y,B):=\exists Z.\mathrm{frac}(X,Y,Z,B)\label{eq:qeq}
\end{equation}
holds if and only if $X,Y$ are uniform, and $B\sim\mathrm{Bern}(|\mathcal{X}|/(|\mathcal{X}|+|\mathcal{Y}|))$,
we know that all rational numbers in $[0,1/2]$ are Bernoulli-definable.
We then show some Bernoulli-definable relations.
\begin{lem}
The relations ``$\le$'', ``$<$'' and ``$=$'' over $[0,1/2]$
are Bernoulli-definable in FOTPI, i.e., there is a first-order formula
$\mathrm{ble}(X,Y)$ which holds if and only if $X\sim\mathrm{Bern}(\theta)$
and $Y\sim\mathrm{Bern}(\phi)$ (up to relabelling), where $0\le\theta\le\phi\le1/2$,
and there are first-order formulae $\mathrm{blt}(X,Y)$, $\mathrm{beq}(X,Y)$
which hold if and only if $\theta<\phi$ and $\theta=\phi$ respectively.
\end{lem}
\begin{IEEEproof}
Consider
\begin{align*}
 & \mathrm{qlt}(X,Y,B)\\
 & :=\mathrm{ueq}_{2}(B)\,\vee\,\Big(\mathrm{card}_{=2}(B)\,\wedge\,\exists C,D.\big(\mathrm{qeq}(X,Y,C)\,\wedge\,\mathrm{ueq}_{2}(D)\\
 & \;\;\;\wedge\,\mathrm{smi}(BCD,C)\,\wedge\,\mathrm{smi}(BCD,D)\,\wedge\,\lnot\mathrm{smi}(BCD,B)\big)\Big).
\end{align*}
We will show that if $X,Y$ are uniform random variables with $|\mathcal{X}|=a$,
$|\mathcal{Y}|=b$, $a<b$, then $\mathrm{qlt}(X,Y,B)$ holds if and
only if $B\sim\mathrm{Bern}(\theta)$ (up to relabelling) for some
$a/(a+b)<\theta\le1/2$. For the ``if'' direction, if $\theta<1/2$
(the case $\theta=1/2$ is clear), take 
\[
(B,C,D)=\begin{cases}
(1,1,0) & \text{with prob.}\;a/(a+b)\\
(1,0,0) & \text{with prob.}\;\theta-a/(a+b)\\
(0,0,0) & \text{with prob.}\;1/2-\theta\\
(0,0,1) & \text{with prob.}\;1/2.
\end{cases}
\]
For the ``only if'' direction, assume $B,C,D\in\{0,1\}$. Assume
the single-mass indicator in $\mathrm{smi}(BCD,C)$ is $\mathbf{1}_{\{1\}}(C)$,
and the indicator in $\mathrm{smi}(BCD,D)$ is $\mathbf{1}_{\{1\}}(D)$.
We must have $\mathbf{P}(C=1)=a/(a+b)$ (if $\mathbf{P}(C=1)=b/(a+b)>1/2$,
then $\mathbf{1}_{\{1\}}(C)$ cannot be a single-mass indicator of
$BCD$). Consider the distribution of $B$. Since we cannot break
the split the masses $\mathbf{P}(C=1)=a/(a+b)$ and $\mathbf{P}(D=1)=1/2$
among different values of $B$, we either assign them to the same
or to different values of $B$. The former case is impossible due
to $\lnot\mathrm{smi}(BCD,B)$. Therefore the masses $a/(a+b)$ and
$1/2$ are assigned to different values of $B$, giving $a/(a+b)\le\gamma\le1/2$.
Note that $\gamma=a/(a+b)$ and $\gamma=1/2$ are impossible due to
$\lnot\mathrm{smi}(BCD,B)$.

We also define
\begin{equation}
\mathrm{qle}(X,Y,B):=\mathrm{qlt}(X,Y,B)\,\vee\,\mathrm{qeq}(X,Y,B).\label{eq:qle}
\end{equation}

Using the fact that for $\theta_{1},\theta_{2}\ge0$, we have $\theta_{1}\le\theta_{2}$
$\Leftrightarrow$ $\forall a,b\in\mathbb{N}_{+}.\,a/b<\theta_{1}\,\to\,a/b<\theta_{2}$,
we can define $\mathrm{ble}(B,C)$, $\mathrm{blt}(B,C)$ and $\mathrm{beq}(B,C)$
by
\begin{align*}
\mathrm{ble}(B,C) & :=\mathrm{card}_{\le2}(B)\,\wedge\,\mathrm{card}_{\le2}(C)\,\\
 & \;\;\;\;\wedge\,\forall X,Y.\big(\mathrm{ult}(X,Y)\,\wedge\,\mathrm{qlt}(X,Y,B)\,\to\,\mathrm{qlt}(X,Y,C)\big),
\end{align*}
\begin{align*}
\mathrm{blt}(B,C) & :=\lnot\mathrm{ble}(C,B),
\end{align*}
\begin{align*}
\mathrm{beq}(B,C) & :=\mathrm{ble}(B,C)\,\wedge\,\mathrm{ble}(C,B).
\end{align*}
\end{IEEEproof}
We can use this to show that any arithmetically definable number \cite{rogers1967theory,boolos2002computability}
in $[0,1/2]$ (i.e., a real number $\theta\in[0,1/2]$ such that the
set $\{(a,b)\in\mathbb{N}_{+}^{2}:\,a/b\le\theta\}$ is definable
using a formula in first-order arithmetic) is Bernoulli-definable.
\begin{thm}
\label{lem:binary_def-1}Any arithmetically definable number in $[0,1/2]$
is Bernoulli-definable in FOTPI.
\end{thm}
\begin{IEEEproof}
Let $\theta\in[0,1/2]$ be an arithmetically definable number. By
Theorem \ref{thm:interpret}, we can find a first-order formula (in
the theory of probabilistic independence) $\psi(X,Y)$ which holds
if and only if $X,Y$ are uniform and $|\mathcal{X}|/|\mathcal{Y}|\le\theta$.
We can check whether $B\sim\mathrm{Bern}(\theta)$ by checking
\[
\forall X,Y.\big(\psi(X,Y)\leftrightarrow\forall C.(\mathrm{qeq}(X,Y,C)\to\mathrm{ble}(C,B))\big).
\]
\end{IEEEproof}
\medskip{}

We call a function $f:\mathbb{N}_{+}\to\mathbb{R}_{\ge0}$ arithmetically
definable if the set $\{(x,a,b)\in\mathbb{N}_{+}^{3}:\,a/b\le f(x)\}$
is definable using a formula in first-order arithmetic. We show that
any arithmetically definable probability mass function is definable
in FOTPI.
\begin{thm}
\label{lem:arith_p}For any probability mass function $p$ over $\mathbb{N}_{+}$,
if the function $p:\mathbb{N}_{+}\to[0,1]$ is arithmetically definable,
then it is definable in FOTPI.
\end{thm}
\begin{IEEEproof}
Let $X\stackrel{\iota}{\le}Y$ where there are at least $3$ possible
values of $Y$ given any $X=x$. We say that two single-mass indicators
$B=\mathbf{1}_{\{b\}}(Y)$, $C=\mathbf{1}_{\{c\}}(Y)$ correspond
to the same value of $X$ if the value of $X$ given $Y=b$ is the
same as the value of $X$ given $Y=c$. This can be checked by
\begin{align*}
\mathrm{smis}(X,Y,B,C) & :=X\stackrel{\iota}{\le}Y\,\wedge\,\mathrm{smi}(Y,B)\,\wedge\,\mathrm{smi}(Y,C)\,\wedge\Big(B\stackrel{\iota}{=}C\,\vee\,\\
 & \;\;\;\;\;\;\,\lnot\exists U.(\mathrm{card}_{\le2}(U)\,\wedge\,BC\stackrel{\iota}{\le}XU)\Big).
\end{align*}
To check this, note that if $B\stackrel{\iota}{\neq}C$ correspond
to the same value of $X$, since there are at least $3$ possible
values of $Y$ corresponding to that value of $X$, and $B,C$ correspond
to $2$ of them, it is impossible to have $\mathrm{card}_{\le2}(U)\,\wedge\,BC\stackrel{\iota}{\le}XU$.
For the other direction, if $B\stackrel{\iota}{\neq}C$ correspond
to different values of $X$, we can take $U=\max\{B,C\}$, and have
$BC\stackrel{\iota}{\le}XU$. We also define the formula for checking
whether $B,C$ correspond to different values of $X$:
\begin{align*}
\mathrm{smid}(X,Y,B,C) & :=X\stackrel{\iota}{\le}Y\,\wedge\,\mathrm{smi}(Y,B)\,\wedge\,\mathrm{smi}(Y,C)\\
 & \;\;\;\;\;\;\,\wedge\,\lnot\mathrm{smis}(X,Y,B,C).
\end{align*}

In order to check whether a random variable $A$ follows $p$, we
assign labels in $\{3,4,\ldots\}$ to values of $A$. Assume $A$
takes values over $\{3,4,\ldots\}$. We call a random variable $L$
a \emph{label} of $A$ if the conditional distribution of $L$ given
$A=a$ is uniform among $a$ different values, and these values are
different for different $a$. This can be checked by (up to relabelling)
\begin{align}
\mathrm{label}_{3}(A,L) & :=A\stackrel{\iota}{\le}L\nonumber \\
 & \;\;\;\wedge\forall B.\big(\mathrm{smi}(L,B)\,\to\,\exists U.\big(\mathrm{uge}_{3}(U)\,\wedge\,U\perp\!\!\!\perp A\,\nonumber \\
 & \;\;\;\;\;\;\wedge\,\forall C.(\mathrm{smis}(A,L,B,C)\to\mathrm{smi}(AU,C))\label{eq:label_smis}\\
 & \;\;\;\;\;\;\wedge\,\forall D.\big(\mathrm{smid}(A,L,B,D)\to\lnot\exists V.(\mathrm{ueq}(U,V)\,\wedge\,V\perp\!\!\!\perp A\,\wedge\,\mathrm{smi}(AV,D))\big)\big)\big).\label{eq:label_smid}
\end{align}
Assume $L$ is a label of $A$. Consider $B$ where $\mathrm{smi}(L,B)$
holds, and assume $B=\mathbf{1}_{\{l\}}(L)$. Let the value of $A$
conditional on $L=l$ be $a$. Then $l$ is one of the $a\ge3$ values
of $L$ corresponding to $A=a$. The line \eqref{eq:label_smis} holds
since we can have $U\sim\mathrm{Unif}[a]$, and any single-mass indicator
$C$ of $L$ corresponding to $A=a$ (there are $a$ such $C$'s)
are single-mass indicators of $AU$ (since $U$ divides the mass $A=a$
into $a$ equal pieces). For \eqref{eq:label_smid}, if $D$ is a
single-mass indicator of $L$ corresponding to a value of $A$ other
than $a$ (let it be $\tilde{a}$), then $\mathbf{P}(D=1)=\mathbf{P}(A=\tilde{a})/\tilde{a}\neq\mathbf{P}(A=\tilde{a})/a$,
and it is impossible to have $\mathrm{ueq}(U,V)\,\wedge\,V\perp\!\!\!\perp A\,\wedge\,\mathrm{smi}(AV,D)$.
For the other direction, using similar arguments, we can deduce that
if $\mathrm{label}_{3}(A,L)$ holds, then conditional on any $A=a$,
$L$ is uniformly distributed in a set $S_{a}$ (of size that equals
the size of $U$ in the definition of $\mathrm{label}_{3}(A,L)$),
and the sizes of $S_{a}$ are distinct (due to \eqref{eq:label_smid}),
and hence we can assign the labels $a=|S_{a}|$ to the values of $A$.

Given $A$ with label $L$, we call $B$ a \emph{divided mass} of
the value $A=|\mathcal{U}|$ if $B=\mathbf{1}_{\{l\}}(L)$ is a single-mass
indicator of $L$, and we have $A=|\mathcal{U}|$ given $L=l$. Note
that this implies $\mathbf{P}(B=1)=\mathbf{P}(A=|\mathcal{U}|)/|\mathcal{U}|$.
This can be checked by
\begin{align}
 & \mathrm{divmass}_{3}(A,L,U,B)\nonumber \\
 & :=\exists\tilde{U}.\big(\mathrm{label}_{3}(A,L)\,\wedge\,\mathrm{smi}(L,B)\,\wedge\,\mathrm{ueq}(U,\tilde{U})\,\nonumber \\
 & \;\;\;\;\wedge\,\mathrm{uge}_{3}(U)\,\wedge\,\tilde{U}\perp\!\!\!\perp A\,\wedge\,\mathrm{smi}(A\tilde{U},B)\big).\label{eq:divmass}
\end{align}

Let $p$ be an arithmetically definable probability mass function
over $\{3,4,\ldots\}$ (we can use the domain $\{3,4,\ldots\}$ instead
of $\mathbb{N}_{+}$ by shifting). Let $\tilde{p}:\{3,4,\ldots\}\to[0,1/3]$
be defined as $\tilde{p}(a):=p(a)/a$ (which is also arithmetically
definable). By Theorem \ref{thm:interpret}, we can find a first-order
formula (in the theory of probabilistic independence) $\psi(W,X,Y)$
which holds if and only if $W,X,Y$ are uniform, $|\mathcal{W}|\ge3$
and $|\mathcal{X}|/(|\mathcal{X}|+|\mathcal{Y}|)\le\tilde{p}(|\mathcal{W}|)$.
To show that $p$ is definable in $\mathcal{L}$, we can check whether
$A\sim p$ (up to relabelling) by
\begin{align*}
 & \exists L.\Big(\mathrm{label}_{3}(A,L)\,\wedge\,\forall B,U.\Big(\mathrm{divmass}_{3}(A,L,U,B)\\
 & \;\;\to\,\forall X,Y,C.\big(\mathrm{ult}(X,Y)\,\wedge\,\mathrm{qeq}(X,Y,C)\,\to\,(\mathrm{ble}(C,B)\leftrightarrow\psi(U,X,Y))\big)\Big)\Big).
\end{align*}
By $\mathrm{divmass}_{3}$, the $B$ (assume $B\sim\mathrm{Bern}(\theta)$,
$\theta\le1/2$) and $U$ in the above definition satisfies $\mathbf{P}(A=|\mathcal{U}|)=\theta|\mathcal{U}|$.
The second line of the definition states that for any uniform $X,Y$
with $|\mathcal{X}|<|\mathcal{Y}|$ and $C\sim\mathrm{Bern}(|\mathcal{X}|/(|\mathcal{X}|+|\mathcal{Y}|))$,
we have $|\mathcal{X}|/(|\mathcal{X}|+|\mathcal{Y}|)\le\theta=\mathbf{P}(A=|\mathcal{U}|)/|\mathcal{U}|$
if and only if $\psi(U,X,Y)\Leftrightarrow|\mathcal{X}|/(|\mathcal{X}|+|\mathcal{Y}|)\le\tilde{p}(|\mathcal{U}|)=p(|\mathcal{U}|)/|\mathcal{U}|$.
\end{IEEEproof}
\medskip{}

We say that $X,Y$ have the same distribution up to relabelling, written
as $X\stackrel{r}{=}Y$, if there exists an injective function $f$
such that $f(X)$ has the same distribution as $Y$. This relation
is also definable in FOTPI.
\begin{prop}
\label{prop:samedist}The following relations over random variables
are definable in FOTPI:
\begin{enumerate}
\item The ``same distribution up to relabelling'' relation $X\stackrel{r}{=}Y$.
\item Comparison of cardinality: $|\mathcal{X}|=|\mathcal{Y}|$ and $|\mathcal{X}|\le|\mathcal{Y}|$.
\end{enumerate}
\end{prop}
\begin{IEEEproof}
Using similar arguments as in Theorem \ref{lem:arith_p}, we can check
whether $A_{1}\stackrel{r}{=}A_{2}$ by
\begin{align*}
 & \exists L_{1},L_{2}.\Big(\mathrm{label}_{3}(A_{1},L_{1})\,\wedge\,\mathrm{label}_{3}(A_{2},L_{2})\,\\
 & \;\wedge\,\forall B,U.\big((\exists B_{1},U_{1}.(\mathrm{beq}(B,B_{1})\,\wedge\,\mathrm{ueq}(U,U_{1})\,\wedge\,\mathrm{divmass}_{3}(A_{1},L_{1},U_{1},B_{1})))\\
 & \;\;\;\;\leftrightarrow(\exists B_{2},U_{2}.(\mathrm{beq}(B,B_{2})\,\wedge\,\mathrm{ueq}(U,U_{2})\,\wedge\,\mathrm{divmass}_{3}(A_{2},L_{2},U_{2},B_{2})))\big)\Big).
\end{align*}
Intuitively, this means that there is a labelling of $A_{1},A_{2}$
such that for any $B\sim\mathrm{Bern}(\theta)$ and uniform $U$,
we have $\mathbf{P}(A_{1}=|\mathcal{U}|)=\theta|\mathcal{U}|$ if
and only if $\mathbf{P}(A_{2}=|\mathcal{U}|)=\theta|\mathcal{U}|$,
which clearly implies $A_{1}$ has the same distribution as $A_{2}$.

We can check whether $V$ is uniform and $|\mathcal{A}|+2\le|\mathcal{V}|$
by
\begin{align*}
\mathrm{cardleu2}(A,V) & :=\exists L.\big(\mathrm{label}_{3}(A,L)\,\\
 & \;\;\;\wedge\,\forall B,U.\big(\mathrm{divmass}_{3}(A,L,U,B)\,\to\,\mathrm{ule}(U,V)\big)\big).
\end{align*}
The reason is that the smallest possible $\max\mathcal{A}$ among
labellings of $A$ using the set of values $\{3,4,\ldots\}$ is $|\mathcal{A}|+2$.
We can then check for $|\mathcal{A}_{1}|\le|\mathcal{A}_{2}|$ by
\[
\forall V.\big(\mathrm{cardleu2}(A_{2},V)\to\mathrm{cardleu2}(A_{1},V)\big),
\]
and we can check for $|\mathcal{A}_{1}|=|\mathcal{A}_{2}|$ by
\[
\forall V.\big(\mathrm{cardleu2}(A_{1},V)\leftrightarrow\mathrm{cardleu2}(A_{2},V)\big).
\]
\end{IEEEproof}
\medskip{}

\section{Representation of Events\label{sec:event}}

In this section, we discuss a representation of events. While the
event $E$ can be represented by the indicator random variable $C=\mathbf{1}\{E\}$,
there is an ambiguity since $C$ can also be the representation of
the complement $E^{c}$ (we do not concern the labelling of $C$). 

Instead, we represent an event $E$ with $\mathbf{P}(E)<1$ as a random
variable $D$ where $D\sim\mathrm{Unif}[k]$ conditional on $E$,
where $k\ge2$ satisfies $\mathbf{P}(E)/k<\mathbf{P}(E^{c})$, and
$D=0$ if $E$ does not occur ($E$ can be recovered by taking the
complement of the largest mass of $D$). If $\mathbf{P}(E)=1$, it
is represented by any $D\sim\mathrm{Unif}[k]$ where $k\ge2$. Note
that there are only two cases where $D$ is uniform: $\mathbf{P}(E)=1$
(which can be checked by $\mathrm{uge}_{2}(D)$) and $\mathbf{P}(E)=0$
(which can be checked by $D\stackrel{\iota}{=}\emptyset$). Technically,
$D$ represents $E$ only up to a difference of a set of measure $0$,
though measure $0$ sets do not affect the truth value of formulae
concerning probabilistic independence.

We can check whether $C$ is the indicator function of the event represented
by $D$ using
\begin{align*}
\mathrm{ind}(D,C) & :=(\mathrm{unif}(D)\,\wedge\,C\stackrel{\iota}{=}\emptyset)\\
 & \;\;\;\vee\,\big(\mathrm{card}_{=2}(C)\,\wedge\,\mathrm{smi}(D,C)\,\wedge\,\exists U,V.\big(\\
 & \;\;\;\;\;\mathrm{uge}_{2}(U)\,\wedge\,\mathrm{ueq}_{2}(V)\,\wedge\,U\perp\!\!\!\perp V\perp\!\!\!\perp C\,\\
 & \;\;\;\;\;\wedge\,D\stackrel{\iota}{\le}CU\,\wedge\,|\mathcal{D}|=|\mathcal{U}|+1\,\\
 & \;\;\;\;\;\wedge\,\forall F,G.\big((\mathrm{smi}(DV,F)\,\wedge\,\lnot\mathrm{smi}(CV,F)\,\wedge\,\mathrm{smi}(DV,G)\,\wedge\,\mathrm{smi}(CV,G))\\
 & \;\;\;\;\;\;\;\to\,\mathrm{blt}(F,G)\big)\big)\big)
\end{align*}
Note that $|\mathcal{D}|=|\mathcal{U}|+1$ can be checked using Proposition
\ref{prop:samedist} and Theorem \ref{thm:interpret}. To check the
above formula, note that $\mathrm{card}_{=2}(C)$, $\mathrm{smi}(D,C)$,
$D\stackrel{\iota}{\le}CU$ and $|\mathcal{D}|=|\mathcal{U}|+1$ ensures
that $C,D$ is in the form $C\in\{0,1\}$, $D=0$ if $C=0$, and $D|\{C=1\}\sim\mathrm{Unif}[k]$
(up to relabelling). In the last two lines, note that $DV$ divides
each mass of $D$ into two equal halves, $F$ is a single mass indicator
of $DV$ where $D\neq0$ (and hence $\mathbf{P}(F=1)=\mathbf{P}(C=1)/(2k)$),
and $G$ is a single mass indicator of $DV$ where $D=0$ (and hence
$\mathbf{P}(G=1)=\mathbf{P}(C=0)/2$). The condition $\mathrm{blt}(F,G)$
means $\mathbf{P}(C=1)/(2k)<\mathbf{P}(C=0)/2$, which is the condition
needed for $D$ to be the representation of the event $C=1$. Note
that $V$ is needed since $\mathrm{blt}$ is defined only for Bernoulli
random variables with parameters in $[0,1/2]$.

We can check whether $D$ is the representation of some event by
\begin{align}
\mathrm{isev}(D) & :=\exists C.\,\mathrm{ind}(D,C).\label{eq:isev}
\end{align}
To check whether the event represented by $D_{1}$ is the complement
of the event represented by $D_{2}$ (up to a difference of measure
$0$):
\begin{align*}
\mathrm{compl}(D_{1},D_{2}) & :=\exists C.\big(\mathrm{ind}(D_{1},C)\,\wedge\,\mathrm{ind}(D_{2},C)\,\wedge\,\lnot\mathrm{smi}(D_{1}D_{2},C)\big)\\
 & \;\;\;\wedge\,(\mathrm{uge}_{2}(D_{2})\,\to\,D_{1}\stackrel{\iota}{=}\emptyset)\,\wedge\,(\mathrm{uge}_{2}(D_{1})\,\to\,D_{2}\stackrel{\iota}{=}\emptyset).
\end{align*}
To check whether the event represented by $D_{1}$ is the same as
the event represented by $D_{2}$ (up to a difference of measure $0$):
\begin{align*}
\mathrm{eveq}(D_{1},D_{2}) & :=\exists C.\big(\mathrm{ind}(D_{1},C)\,\wedge\,\mathrm{ind}(D_{2},C)\big)\,\wedge\,\lnot\mathrm{compl}(D_{1},D_{2}).
\end{align*}
To check whether the event represented by $D_{1}$ is a subset of
the event represented by $D_{2}$ (up to a difference of measure $0$):
\begin{align*}
 & \mathrm{subset}(D_{1},D_{2})\\
 & :=\mathrm{isev}(D_{1})\,\wedge\,\mathrm{isev}(D_{2})\\
 & \;\;\wedge\Big(D_{1}\stackrel{\iota}{=}\emptyset\,\vee\,\mathrm{uge}_{2}(D_{2})\vee\,\big(\lnot\mathrm{unif}(D_{1})\,\wedge\,\lnot\mathrm{unif}(D_{2})\,\wedge\\
 & \;\;\;\;\;\;\;\exists C_{2}.\big(\mathrm{ind}(D_{2},C_{2})\,\wedge\,\forall\tilde{D}_{1}.(\mathrm{eveq}(D_{1},\tilde{D}_{1})\to\mathrm{smi}(D_{2}\tilde{D}_{1},C_{2}))\big)\big)\Big).
\end{align*}
The reason is that if the nondegenerate event represented by $D_{1}$
(let it be $E_{1}$) is a subset of the nondegenerate event represented
by $D_{2}$ (let it be $E_{2}$), then any representation $\tilde{D}_{1}$
of $E_{1}$ will be constant conditional on $E_{2}^{c}$, and hence
$C_{2}=\mathbf{1}\{E_{2}\}$ is a single mass indicator of $\tilde{D}_{1}$.
For the other direction, if $\mathbf{P}(E_{1}\backslash E_{2})>0$,
then we can have $\tilde{D}_{1}|E_{1}\sim\mathrm{Unif}[k]$ for $k$
large enough so that $\mathbf{P}(E_{1})/k<\mathbf{P}(E_{1}\backslash E_{2})$
, and hence $\tilde{D}_{1}$ is not constant conditional on $E_{1}\backslash E_{2}$,
and hence is not constant conditional on $E_{2}^{c}$. This means
$\mathrm{smi}(D_{2}\tilde{D}_{1},C_{2})$ cannot hold.

We can also take the union of a collection of events. Let $P$ be
a first-order formula. To check whether $D$ is the representation
of the union of all events $E$ with a representation satisfying $P$:
\begin{align}
\mathrm{union}_{P}(D) & :=\forall D_{2}.\big(P(D_{2})\,\wedge\,\mathrm{isev}(D_{2})\,\to\,\mathrm{subset}(D_{2},D)\big)\nonumber \\
 & \;\;\;\wedge\forall\tilde{D}.\Big(\mathrm{isev}(\tilde{D})\,\wedge\,\forall D_{2}.\big(P(D_{2})\,\wedge\,\mathrm{isev}(D_{2})\,\to\,\mathrm{subset}(D_{2},\tilde{D})\big)\nonumber \\
 & \;\;\;\;\;\;\;\;\;\;\;\to\mathrm{subset}(D,\tilde{D})\Big).\label{eq:union}
\end{align}
Technically, since a representation only identifies the event up to
a difference of measure $0$, an uncountable union may not be well-defined
(in the equivalence classes of events mod $0$). Instead of the ordinary
union of sets, the above definition actually describes the essential
union of measurable sets \cite[Def. 2]{pales2020essential}, which
is always measurable. Nevertheless, since this paper concerns discrete
settings, we can regard essential union as ordinary union. We also
define
\begin{equation}
\mathrm{union}(D_{1},\ldots,D_{n},\tilde{D}):=\mathrm{union}_{D:\bigvee_{i}\mathrm{eveq}(D,D_{i})}(\tilde{D}).\label{eq:union_finite}
\end{equation}
Define $\mathrm{inter}$ for intersection similarly.

We can also check whether the event represented by $D_{1}$ is disjoint
of the event represented by $D_{2}$ (up to a difference of measure
$0$):
\begin{align*}
 & \mathrm{disjoint}(D_{1},D_{2})\\
 & :=\exists\tilde{D}_{1}.\big(\mathrm{compl}(D_{1},\tilde{D}_{1})\,\wedge\,\mathrm{subset}(D_{2},\tilde{D}_{1})\big).
\end{align*}
To check whether the event represented by $D_{1}$ is independent
of the event represented by $D_{2}$:
\begin{align*}
 & \mathrm{indep}(D_{1},D_{2})\\
 & :=\exists\tilde{D}_{1},\tilde{D}_{2}.\big(\mathrm{eveq}(D_{1},\tilde{D}_{1})\,\wedge\,\mathrm{eveq}(D_{2},\tilde{D}_{2})\,\wedge\,\tilde{D}_{1}\perp\!\!\!\perp\tilde{D}_{2}\big).
\end{align*}

To check whether $\mathbf{P}(E_{1})\le\mathbf{P}(E_{2})$, where $E_{i}$
is represented by $D_{i}$:
\begin{equation}
\mathrm{prle}(D_{1},D_{2}):=\exists\tilde{D}_{1}.\big(D_{1}\stackrel{r}{=}\tilde{D}_{1}\,\wedge\,\mathrm{subset}(\tilde{D}_{1},D_{2})\big).\label{eq:prle}
\end{equation}
Also define
\begin{equation}
\mathrm{preq}(D_{1},D_{2}):=\exists\tilde{D}_{1}.\big(D_{1}\stackrel{r}{=}\tilde{D}_{1}\,\wedge\,\mathrm{eveq}(\tilde{D}_{1},D_{2})\big).\label{eq:preq}
\end{equation}
Note that this allows us to perform addition and multiplication on
probability of events. For example, to check whether $\mathbf{P}(E_{1})=\mathbf{P}(E_{2})\mathbf{P}(E_{3})+\mathbf{P}(E_{4})$:
\begin{align}
 & \exists\tilde{D}_{1},\ldots,\tilde{D}_{4},\tilde{D}_{23}.\Big(\bigwedge_{i=1}^{4}\mathrm{preq}(D_{i},\tilde{D}_{i})\nonumber \\
 & \;\;\wedge\,\mathrm{indep}(\tilde{D}_{2},\tilde{D}_{3})\,\wedge\,\mathrm{inter}(\tilde{D}_{2},\tilde{D}_{3},\tilde{D}_{23})\nonumber \\
 & \;\;\wedge\,\mathrm{disjoint}(\tilde{D}_{23},\tilde{D}_{4})\,\wedge\,\mathrm{union}(\tilde{D}_{23},\tilde{D}_{4},\tilde{D}_{1})\Big).\label{eq:evpr_eg}
\end{align}

Given $A$ with label $L$ \eqref{eq:label_smid}, we can check whether
$\mathbf{P}(A=|\mathcal{U}|)>0$, and $D$ is the representation of
the event $A=|\mathcal{U}|$ (where $U$ is a uniform random variable)
by
\begin{align*}
 & \mathrm{labelevne}_{3}(A,L,U,D)\\
 & :=\exists C,\tilde{U}.\Big(\mathrm{ind}(D,C)\,\wedge\,\mathrm{label}_{3}(A,L)\,\wedge\,\mathrm{smi}(A,C)\,\wedge\,\mathrm{ueq}(U,\tilde{U})\,\\
 & \;\;\;\;\wedge\,\mathrm{uge}_{3}(U)\,\wedge\,\tilde{U}\perp\!\!\!\perp A\,\\
 & \;\;\;\;\wedge\,\forall B.\big(\mathrm{divmass}_{3}(A,L,U,B)\,\to\,\mathrm{smi}(A\tilde{U},B)\,\wedge\,\mathrm{smi}(C\tilde{U},B)\\
 & \;\;\;\;\;\;\;\;\wedge\,\exists\tilde{D}.(\mathrm{eveq}(D,\tilde{D})\,\wedge\,B\stackrel{\iota}{\le}\tilde{D})\big)\Big).
\end{align*}
To show the validity of this formula, we let $C=\mathbf{1}_{\{a\}}(A)$.
Note that if $|\mathcal{A}|\ge3$ (otherwise the above formula is
obviously valid), $\mathrm{smi}(A\tilde{U},B)\,\wedge\,\mathrm{smi}(C\tilde{U},B)$
implies that $B=1$ (assuming $B=\mathbf{1}_{\{l\}}(L)$) only if
$C=1$. Hence $\mathrm{smi}(C\tilde{U},B)$ implies $\mathbf{P}(B=1)=\mathbf{P}(A=a)/|\mathcal{U}|$,
and the mass $A=a$ is divided into $|\mathcal{U}|$ equal pieces
by $L$, and hence $a=|\mathcal{U}|$ by the definition of $\mathrm{label}_{3}(A,L)$.
It is left to check that $D$ is the representation of the event $A=|\mathcal{U}|$
(instead of $A\neq|\mathcal{U}|$). This is ensured by $\exists\tilde{D}.(\mathrm{eveq}(D,\tilde{D})\,\wedge\,B\stackrel{\iota}{\le}\tilde{D})$.
If $D$ represents $A\neq|\mathcal{U}|$, then $\tilde{D}$ is constant
given $A=|\mathcal{U}|$, so $B$ cannot be a function of $\tilde{D}$
since $\mathbf{P}(B=1)=\mathbf{P}(A=|\mathcal{U}|)/|\mathcal{U}|$.
If $D$ represents $A=|\mathcal{U}|$, we can take $\tilde{D}|\{A=|\mathcal{U}|\}\sim\mathrm{Unif}[k|\mathcal{U}|]$
(for large enough $k$ so $\tilde{D}$ satisfies the definition of
a representation of an event) such that $L$ is a function of $\tilde{D}$
conditional on $A=|\mathcal{U}|$.

Note that $\mathrm{labelevne}_{3}$ is false if $\mathbf{P}(A=|\mathcal{U}|)=0$.
To check whether $D$ is the representation of the event $A=|\mathcal{U}|$
(which is empty if $\mathbf{P}(A=|\mathcal{U}|)=0$), we use
\begin{align*}
 & \mathrm{labelev}_{3}(A,L,U,D)\\
 & :=\mathrm{labelevne}_{3}(A,L,U,D)\\
 & \;\;\;\vee\big(D\stackrel{\iota}{=}\emptyset\,\wedge\,\lnot\exists D_{2}.\mathrm{labelevne}_{3}(A,L,U,D_{2})\big).
\end{align*}

The condition that $X$ has the same distribution as $Z$ is written
as $X\stackrel{d}{=}Z$. We say that $Y|X$ follows the conditional
distribution of $W|Z$, written as $Y|X\sim W|Z$, if $p_{X}(x)>0$
implies $p_{Z}(x)>0$ and $p_{Y|X}(y|x)=p_{W|Z}(y|x)$ for any $y$.
Note that since FOTPI does not concern the labelling of a random variable,
we have to use another random variable $L_{X}$ (the label) to specify
the values of $X$, as described in \eqref{eq:label_smid}. The statements
$X\stackrel{d}{=}Z$ and $Y|X\sim W|Z$ are not valid statements in
FOTPI without the labels. 

Also, we say that $Y|X$ follows the conditional distribution of $W|Z$
up to relabelling, written as $Y|X\stackrel{r}{\sim}W|Z$, if there
exists relabellings $\tilde{X}\stackrel{\iota}{=}X$, $\tilde{Y}\stackrel{\iota}{=}Y$,
$\tilde{Z}\stackrel{\iota}{=}Z$, $\tilde{W}\stackrel{\iota}{=}W$
such that $\tilde{Y}|\tilde{X}\sim\tilde{W}|\tilde{Z}$. Note that
$Y|X\stackrel{r}{\sim}W|Z$ does not depend on the labelling of $X,Y,Z,W$.
\begin{prop}
\label{prop:samecd}The following conditions are definable in FOTPI:
\begin{itemize}
\item The condition $X\stackrel{d}{=}Z$, where $L_{X},L_{Z}$ are the labels
of $X,Z$ respectively \eqref{eq:label_smid}.
\item The condition $Y|X\sim W|Z$, where $L_{X},L_{Y},L_{Z},L_{W}$ are
the labels of $X,Y,Z,W$ respectively \eqref{eq:label_smid}.
\item The condition $Y|X\stackrel{r}{\sim}W|Z$.
\end{itemize}
\end{prop}
\begin{IEEEproof}
Note that $X\stackrel{d}{=}Z$ can be checked by
\begin{align*}
 & \mathrm{deq}(X,L_{X},Z,L_{Z})\\
 & :=\mathrm{label}_{3}(X,L_{X})\,\wedge\,\mathrm{label}_{3}(Z,L_{Z})\,\wedge\,\forall U,D_{X},D_{Z}.\big(\\
 & \;\;\;\;\;\mathrm{labelev}_{3}(X,L_{X},U,D_{X})\,\wedge\,\mathrm{labelev}_{3}(Z,L_{Z},U,D_{Z})\\
 & \;\;\;\;\;\to\,\mathbf{P}(D_{X})=\mathbf{P}(D_{Z})\big),
\end{align*}
where $\mathbf{P}(D_{X})=\mathbf{P}(D_{Z})$ is checked by \eqref{eq:preq}.
Also, $Y|X\sim W|Z$ can be checked by
\begin{align*}
 & \mathrm{cdeq}(X,L_{X},Y,L_{Y},Z,L_{Z},W,L_{W})\\
 & :=\mathrm{label}_{3}(X,L_{X})\,\wedge\,\mathrm{label}_{3}(Y,L_{Y})\,\wedge\,\mathrm{label}_{3}(Z,L_{Z})\,\wedge\,\mathrm{label}_{3}(W,L_{W})\\
 & \;\;\wedge\,\forall U,V,D_{X},D_{Y},D_{Z},D_{W}.\big(\\
 & \;\;\;\;\;D_{X}\stackrel{\iota}{\neq}\emptyset\,\wedge\,\mathrm{labelev}_{3}(X,L_{X},U,D_{X})\,\wedge\,\mathrm{labelev}_{3}(Z,L_{Z},U,D_{Z})\\
 & \;\;\;\;\;\wedge\,\mathrm{labelev}_{3}(Y,L_{Y},V,D_{Y})\,\wedge\,\mathrm{labelev}_{3}(W,L_{W},V,D_{W})\\
 & \;\;\;\;\;\to\,D_{Z}\stackrel{\iota}{\neq}\emptyset\,\wedge\,\mathbf{P}(D_{X})\mathbf{P}(D_{Z}\cap D_{W})=\mathbf{P}(D_{Z})\mathbf{P}(D_{X}\cap D_{Y})\big),
\end{align*}
where $D_{Z}\cap D_{W}$ denotes the representation of the intersection
of the events represented by $D_{Z}$ and $D_{W}$ \eqref{eq:union_finite},
$\mathbf{P}(D_{X})$ denotes the probability of the event represented
by $D_{X}$, and $\mathbf{P}(D_{X})\mathbf{P}(D_{Z}\cap D_{W})=\mathbf{P}(D_{Z})\mathbf{P}(D_{X}\cap D_{Y})$
can be expressed in the same way as \eqref{eq:evpr_eg}. Note that
$Y|X\sim W|Z$ if and only if $p_{X}(x)>0\,\to\,p_{Z}(x)>0\,\wedge\,p_{X}(x)p_{Z,W}(x,y)=p_{Z}(x)p_{X,Y}(x,y)$
for all $x,y$.

For $Y|X\stackrel{r}{\sim}W|Z$, it can be checked by
\begin{align*}
 & \mathrm{cdeqr}(X,Y,Z,W)\\
 & :=\exists L_{X},L_{Y},L_{Z},L_{W}.\mathrm{cdeq}(X,L_{X},Y,L_{Y},Z,L_{Z},W,L_{W}).
\end{align*}
\end{IEEEproof}
\[
\]

\section{Representation of Random Sequences\label{sec:seq}}

In the remainder of this paper, we use the following notations within
first-order formulae:
\begin{itemize}
\item Random variables are denoted by uppercase letters (except $E$).
\end{itemize}
\smallskip{}

\begin{itemize}
\item $\mathbb{N}_{0}$-valued variables are denoted by lowercase (English
or Greek) letters. Any $\mathbb{N}_{0}$-valued variable $\alpha$
is understood as a random variable obtained using the representation
in Theorem \ref{thm:interpret}, restricted to satisfy $\mathrm{isnat}(\alpha)$
\eqref{eq:isnat}, and hence can be expressed in FOTPI. Addition,
comparison and multiplication can be performed on $\mathbb{N}_{0}$-valued
variables due to Theorem \ref{thm:interpret}.
\end{itemize}
\smallskip{}

\begin{itemize}
\item Event variables are denoted by $E$ (or with subscripts). Any event
variable $E$ is understood as a random variable obtained using the
representation in Section \ref{sec:event}, restricted to satisfy
$\mathrm{isev}(E)$ \eqref{eq:isev}, and hence can be expressed in
FOTPI. Set operations and relations such as $E_{1}\cup E_{2}$, $E_{1}\cap E_{2}$,
$E_{1}\stackrel{as}{=}E_{2}$ (i.e., $E_{1}\leftrightarrow E_{2}$
holds almost surely), $E_{1}\stackrel{as}{\subseteq}E_{2}$ (i.e.,
$E_{1}\to E_{2}$ holds almost surely) and $E_{1}\perp\!\!\!\perp E_{2}$
can be defined (see Section \ref{sec:event}). For the union of a
collection of events $E$ satisfying $P(E)$ \eqref{eq:union}, we
use the notation $\bigcup_{E:P(E)}E$.
\end{itemize}
\smallskip{}

We have defined $\mathrm{label}_{3}(A,L)$ \eqref{eq:label_smid}
to check whether $A$ can be labelled using values in $\{3,4,\ldots\}$,
and $L|A\sim\mathrm{Unif}[A]$. For the purpose of simplicity, we
will shift the values of $A$ to $\{0,1,\ldots\}$ (i.e., $L|A\sim\mathrm{Unif}[A+3]$).
Given $A\in\{0,1,\ldots\}$ with label $L$, we can check whether
$E$ is (the representation of) the event $A=a$, $a\in\mathbb{N}_{0}$
(recall that $a$ is understood to be obtained using the representation
in Theorem \ref{thm:interpret}) by
\[
\mathrm{labelev}_{0}(A,L,a,E):=\mathrm{labelev}_{3}(A,L,a+3,E).
\]
For notational simplicity, we will denote the $E$ satisfying the
above formula by
\begin{equation}
\{A\circ L=a\}.\label{eq:ev_notation}
\end{equation}
The notation $A\circ L$ intuitively means the random variable $A$
with labels given by $L$ (note that $A$ itself, as for any random
variable in FOTPI, does not inherently have well-defined values, since
formulae in FOTPI do not concern the labelling of random variables).

We can also obtain the event where the random variable $A$ (with
label $L$) equals the random variable $B$ (with label $M$) by
\[
\bigcup_{E:\,\exists a.\,E\stackrel{as}{=}\{A\circ L=a\}\cap\{B\circ M=a\}}E,
\]
which is defined using \eqref{eq:union}. The above event is denoted
as
\begin{equation}
\{A\circ L=B\circ M\}.\label{eq:event_notation}
\end{equation}
If the above event occurs with probability $1$ (recall that we can
check whether $\mathbf{P}(E)=1$ by $\mathrm{ueq}_{2}(E)$), we simply
denote this condition as
\[
A\circ L\stackrel{as}{=}B\circ M.
\]
The conditional distribution relation (Proposition \ref{prop:samecd})
is denoted as
\begin{equation}
Y\circ L_{Y}|X\circ L_{X}\sim W\circ L_{W}|Z\circ L_{Z}.\label{eq:samecd_notation}
\end{equation}
When we check independence or conditional independence for labelled
random variables, e.g.,
\begin{equation}
A\circ L\perp\!\!\!\perp B\circ M,\label{eq:label_indep}
\end{equation}
we ignore the labels (the above line means $A\perp\!\!\!\perp B$).

In this section, we use the notation $X^{n}=(X_{1},\ldots,X_{n})$
to denote a sequence of random variables. The main challenge of expressing
a coding setting in FOTPI is that coding settings are often defined
asymptotically, where $n$, the length of the random sequences, tends
to infinity. However, the number of random variables in a formula
in FOTPI is fixed. Therefore, we have to design a method to extract
$X_{i}$, given $X^{n}$ as a single random variable. We utilize the
classical Gödel encoding \cite{godel1931formal,godel1934undecidable,smith2013introduction},
which we briefly recall below:
\begin{defn}
The \emph{Gödel beta function }\cite{godel1934undecidable} is defined
as
\[
beta(b,c,i):=b\,\mathrm{mod}\,(c(i+1)+1),
\]
where $a\,\mathrm{mod}\,b$ denotes the remainder when $a$ is divided
by $b$. It satisfies the property that for any sequence $a_{0},a_{1},\ldots,a_{n}\in\mathbb{N}_{0}$,
there exists $b,c\in\mathbb{N}_{0}$ such that $beta(b,c,i)=a_{i}$
for $i\in[0..n]$. The \emph{Cantor pairing function} (which is a
bijection from $\mathbb{N}_{0}\times\mathbb{N}_{0}$ to $\mathbb{N}_{0}$)
is defined as
\[
pair(b,c):=\frac{(b+c)(b+c+1)}{2}+c.
\]
For a sequence $a_{1},a_{2},\ldots,a_{n}\in\mathbb{N}_{0}$, its \emph{Gödel
encoding} is defined as
\[
enc(\{a_{i}\}_{i\in[n]}):=pair(b,c),
\]
where $b,c\in\mathbb{N}_{0}$ satisfy $beta(b,c,0)=n$ and $beta(b,c,i)=a_{i}$
for $i\in[n]$ (if multiple $(b,c)$ satisfies the requirements, take
the smallest $pair(b,c)$). The decoding predicate $\mathrm{dec}(r,i,a)$
is defined so that $\mathrm{dec}(enc(\{a_{i}\}_{i\in[n]}),i,a)$ is
true if and only if $i\in[n]$ and $a_{i}=a$. It can be defined by
\begin{align*}
\mathrm{decn}(r,i,a) & :=\exists b,c.\big(pair(b,c)=r\,\\
 & \;\;\;\wedge\,1\le i\le beta(b,c,0)\,\wedge\,beta(b,c,i)=a\big),
\end{align*}
\begin{align*}
\mathrm{dec}(r,i,a) & :=\mathrm{decn}(r,i,a)\,\wedge\,\\
 & \;\;\;\forall r'.\big((\forall i',a'.(\mathrm{decn}(r,i',a')\leftrightarrow\mathrm{decn}(r',i',a')))\to r'\ge r\big).
\end{align*}
Note that the definition of $\mathrm{dec}$ enforces the minimality
condition in the definition of $enc$ ($r$ is the smallest among
all $r'$ which gives the same decoded values for all $i$). Due to
Theorem \ref{thm:interpret}, $\mathrm{dec}$ can be defined in FOTPI
(though $enc$ cannot since FOTPI does not natively support quantifying
over sequences).
\end{defn}

Assume $X_{1},\ldots,X_{n}\in\mathbb{N}_{0}$. Let $\bar{X}:=enc(X^{n})$
be the Gödel encoding of $X^{n}$ (which is a random integer), and
$\bar{L}|\bar{X}\sim\mathrm{Unif}[\bar{X}+3]$ is the label of $\bar{X}$.
We can check whether $\bar{X}=enc(X^{n})$ for some $X^{n}$ by the
following formula in FOTPI:
\begin{align*}
\mathrm{isseq}(\bar{X},\bar{L},n) & :=\mathrm{label}_{3}(\bar{X},\bar{L})\,\wedge\,\\
 & \;\;\;\forall l.\big(\{\bar{X}[\bar{L}]=l\}\neq\emptyset\\
 & \;\;\;\;\to\,\big(\forall i.(1\le i\le n\,\to\,\exists x.\,\mathrm{dec}(l,i,x))\big)\\
 & \;\;\;\;\;\;\;\;\;\wedge\,\big(\forall i,x.(i>n\,\to\,\lnot\mathrm{dec}(l,i,x))\big)\big),
\end{align*}
where $\{\bar{X}[\bar{L}]=l\}$ is defined in \eqref{eq:ev_notation}.
Intuitively, this means if $\mathbf{P}(\bar{X}=l)>0$, then we can
decode the $i$-th entry of $l$ for $1\le i\le n$, and we cannot
decode its $i$-th entry for $i>n$. 

We now define a formula to check whether $X$ is the $i$-th component
(i.e., $X_{i}$) of $\bar{X}$. Let $L|X\sim\mathrm{Unif}[X+3]$ be
the labelling of $X$. This can be checked by
\begin{align*}
 & \mathrm{entry}(\bar{X},\bar{L},n,X,L,i)\\
 & :=\mathrm{isseq}(\bar{X},\bar{L},n)\,\wedge\,\mathrm{label}_{3}(X,L)\,\wedge\,1\le i\le n\\
 & \;\;\;\wedge\,\forall x,l.\Big(\mathrm{dec}(l,i,x)\,\to\,\{\bar{X}\circ\bar{L}=l\}\stackrel{as}{\subseteq}\{X\circ L=x\}\Big).
\end{align*}
This means that if $\mathrm{dec}(l,i,x)$, then $\bar{X}=l$ implies
$X=x$.

We can also obtain subsequences of $\bar{X}=enc(X^{n})$. To check
whether $\bar{Y}$ with label $\bar{M}$ satisfies $\bar{Y}=enc(X_{i},\ldots,X_{j})$,
we use the formula
\begin{align*}
 & \mathrm{subseq}(\bar{X},\bar{L},n,\bar{Y},\bar{M},i,j)\\
 & :=1\le i\le j\le n\,\wedge\,\mathrm{isseq}(\bar{X},\bar{L},n)\,\wedge\,\mathrm{isseq}(\bar{Y},\bar{M},j-i+1)\\
 & \;\;\;\wedge\,\forall k.\big(i\le k\le j\,\to\,\\
 & \;\;\;\;\;\;\forall X,L.\big(\mathrm{entry}(\bar{X},\bar{L},n,X,L,k)\,\\
 & \;\;\;\;\;\;\;\;\;\;\;\;\leftrightarrow\,\mathrm{entry}(\bar{Y},\bar{M},j-i+1,X,L,k-i+1)\big)\big).
\end{align*}
For notational simplicity, in the remainder of this paper, we use
the notation
\[
X\circ L=(\bar{X}\circ\bar{L})_{i}
\]
to denote $\mathrm{entry}(\bar{X},\bar{L},X,L,i)$, and
\[
\bar{Y}\circ\bar{M}=(\bar{X}\circ\bar{L})_{i..j}
\]
to denote $\mathrm{subseq}(\bar{X},\bar{L},\bar{Y},\bar{M},i,j)$.
Given $X_{1},\ldots,X_{n}$ with labels $L_{1},\ldots,L_{n}$ for
a fixed (non-variable) $n$, we write
\begin{equation}
\bar{X}\circ\bar{L}=(X_{1}\circ L_{1},\ldots,X_{n}\circ L_{n})\label{eq:pair}
\end{equation}
if 
\[
\mathrm{isseq}(\bar{X},\bar{L},n)\,\wedge\,\bigwedge_{i=1}^{n}\mathrm{entry}(\bar{X},\bar{L},n,X_{i},L_{i},i),
\]
i.e., $\bar{X}=enc(X^{n})$ with label $\bar{L}$.

To check whether $\bar{X}=enc(X^{n})$ (with label $\bar{L}$) where
$X_{1},\ldots,X_{n}$are i.i.d. with the same distribution as $X$
(with label $L$) ,
\begin{align*}
 & \mathrm{iid}(\bar{X},\bar{L},n,X,L)\\
 & :=\mathrm{isseq}(\bar{X},\bar{L},n)\,\wedge\,\forall i,X',L',\bar{Y},\bar{M}.\big(\\
 & \;\;\;X'\circ L'=(\bar{X}\circ\bar{L})_{i}\,\wedge\,\bar{Y}\circ\bar{M}=(\bar{X}\circ\bar{L})_{1..i-1}\\
 & \;\;\;\,\to\,X'\circ L'\stackrel{d}{=}X\circ L\,\wedge\,X'\perp\!\!\!\perp\bar{Y}\big),
\end{align*}
where $X'\circ L'\stackrel{d}{=}X\circ L$ is defined in \eqref{eq:samecd_notation}
(we can let the conditioned random variables be degenerate). The above
formula checks that $X_{i}\stackrel{d}{=}X$ is independent of $X_{1},\ldots,X_{i-1}$. 

We now characterize the entropy $H(X)$. By the source coding theorem,
$H(X)$ is the smallest $R$ such that for any $R'>R$ and $\epsilon>0$,
there exists $n$, $W$, $Y^{n}$ with $Y^{n}\stackrel{\iota}{\le}W\stackrel{\iota}{\le}X^{n}$,
$|\mathcal{W}|\le2^{nR'}$, and $\mathbf{P}(X^{n}\neq Y^{n})\le\epsilon$,
where $X_{1},\ldots,X_{n}$are i.i.d. with the same distribution as
$X$. Therefore, we can check whether $H(X)\le a/b$ for $a,b\in\mathbb{N}_{0}$,
$b>0$ by
\begin{align}
\mathrm{hle}(X,a,b) & :=\forall a',b'.\big(a'b>ab'\,\to\,\nonumber \\
 & \;\;\;\forall E.\big(\mathrm{isev}(E)\,\wedge\,E\stackrel{\iota}{\neq}\emptyset\to\exists\bar{X},\bar{L},\bar{Y},\bar{M},L,W,n.\big(\nonumber \\
 & \;\;\;\;\;\mathrm{iid}(\bar{X},\bar{L},n,X,L)\,\wedge\,\mathrm{isseq}(\bar{Y},\bar{M},n)\,\nonumber \\
 & \;\;\;\;\;\wedge\,\bar{Y}\stackrel{\iota}{\le}W\stackrel{\iota}{\le}\bar{X}\,\wedge\,|\mathcal{W}|^{b'}\le2^{na'}\nonumber \\
 & \;\;\;\;\;\wedge\,\mathrm{prle}(\{\bar{X}\circ\bar{L}\neq\bar{Y}\circ\bar{M}\},E)\big)\big)\big),\label{eq:hle}
\end{align}
where $|\mathcal{W}|^{b'}\le2^{na'}$ can be defined since exponentiation
is definable using a first-order formula over natural numbers (alternatively,
we can construct an i.i.d. sequence $\bar{W}$ by $\mathrm{iid}(W,L_{W},b',\bar{W},\bar{L}_{\bar{W}})$,
which has cardinality $|\mathcal{W}|^{b'}$, and use Proposition \ref{prop:samedist}
to check $|\mathcal{W}|^{b'}\le2^{na'}$), $\{\bar{X}\circ\bar{L}\neq\bar{Y}\circ\bar{M}\}$
represents the event $X^{n}\neq Y^{n}$ (this notation is defined
in \eqref{eq:event_notation}), and $\mathrm{prle}(\{\bar{X}\circ\bar{L}\neq\bar{Y}\circ\bar{M}\},E)$
\eqref{eq:prle} checks that $\mathbf{P}(X^{n}\neq Y^{n})\le\mathbf{P}(E)$.
The above formula means that for any $a',b'$ such that $a'/b'>a/b$,
for any $E$ with $\mathbf{P}(E)>0$, there exists $n$, $W$, $Y^{n}$
with $Y^{n}\stackrel{\iota}{\le}W\stackrel{\iota}{\le}X^{n}$, $|\mathcal{W}|\le2^{na'/b'}$,
and $\mathbf{P}(X^{n}\neq Y^{n})\le\mathbf{P}(E)$, where $X_{1},\ldots,X_{n}$are
i.i.d. with the same distribution as $X$.

Using \eqref{eq:hle}, we can check inequalities among entropy of
random variables. For example, $H(X)\le2H(Y)$ can be checked by
\[
\forall a,b.\big(\mathrm{hle}(Y,a,2b)\to\mathrm{hle}(X,a,b)\big).
\]

\medskip{}

\section{Single-Letter Characterization of Capacity Regions\label{sec:singleletter}}

In this section, we study a general network which encompasses the
discrete memoryless networks in \cite{aref1981information,el1981information,lim2011noisy},
the finite-state Markov channel \cite{goldsmith1996capacity}, and
the Markov network \cite{agarwal2018non}.
\begin{defn}
[Joint source-channel Markov network] We define a network with $k$
terminals as follows. Consider the source distribution $p_{W_{1},\ldots,W_{k}}$,
the channel $p_{Y_{1},\ldots,Y_{k},S'|X_{1},\ldots,X_{k},S}$, input
alphabet $\mathcal{X}_{i}$ (where $X_{i}\in\mathcal{X}_{i}$), initial
state distribution $p_{S}$, and decoding requirement $p_{Z_{1},\ldots,Z_{k}|W_{1},\ldots,W_{k}}$,
where the random variables take values in $\mathbb{N}_{0}$. At the
beginning of the communication scheme, the source $W_{i,1},\ldots,W_{i,n}$
is given to terminal $i$, where $(W_{1,t},\ldots,W_{k,t})\sim p_{W_{1},\ldots,W_{k}}$
i.i.d. across $t\in[n]$. Let $S_{t},X_{i,t},Y_{i,t}$ be the channel
state, the channel input given by terminal $i$, and the channel output
observed by terminal $i$ at time $t\in[n]$ respectively. Let $S_{1}\sim p_{S}$
independent of $\{W_{i,t}\}_{i,t}$. At time $t\in[n]$, terminal
$i$ outputs $X_{i,t}\in\mathcal{X}_{i}$ as a (possibly stochastic)
mapping of $W_{i,1},\ldots,W_{i,n}$, $X_{i,1},\ldots,X_{i,t-1}$
and $Y_{i,1},\ldots,Y_{i,t-1}$ for $i\in[k]$, and then $Y_{1,t},\ldots,Y_{k,t},S_{t+1}$
are generated given $X_{1,t},\ldots,X_{k,t},S_{t}$ following $p_{Y_{1},\ldots,Y_{k},S'|X_{1},\ldots,X_{k},S}$.
At the end, terminal $i$ outputs $\hat{Z}_{i,1},\ldots,\hat{Z}_{i,n}$
as a (possibly stochastic) mapping of $W_{i,1},\ldots,W_{i,n}$, $X_{i,1},\ldots,X_{i,n}$
and $Y_{i,1},\ldots,Y_{i,n}$. The probability of error is defined
as the total variation distance
\[
P_{e}:=d_{\mathrm{TV}}\big(\{W_{1,t},\ldots,W_{k,t},\hat{Z}_{1,t},\ldots,\hat{Z}_{k,t}\}_{t\in[n]},\,(p_{W_{1},\ldots,W_{k}}p_{Z_{1},\ldots,Z_{k}|W_{1},\ldots,W_{k}})^{n}\big),
\]
i.e., this is a strong coordination problem \cite{cuff2010coordination}
where $W_{1,t},\ldots,W_{k,t},\hat{Z}_{1,t},\ldots,\hat{Z}_{k,t}$
should approximately follow $p_{W_{1},\ldots,W_{k}}p_{Z_{1},\ldots,Z_{k}|W_{1},\ldots,W_{k}}$.
We say that this network admits a communication scheme if for any
$\epsilon>0$, there exists $n$ and a communication scheme (encoding
and decoding functions) such that $P_{e}\le\epsilon$.
\end{defn}
\medskip{}
Note that the joint source-channel Markov network encompasses channel
coding settings by letting $M_{1},\ldots,M_{l}$ to be independent
(they are the messages), $W_{i}=\{M_{j}\}_{j\in\mathcal{E}_{i}}$
where $\mathcal{E}_{i}\subseteq[k]$ is the set of messages that terminal
$k$ can access, and $Z_{i}=\{M_{j}\}_{j\in\mathcal{D}_{i}}$ where
$\mathcal{D}_{i}\subseteq[k]$ is the set of messages that terminal
$k$ intends to decode. In this case, $P_{e}$ is the probability
that any terminal makes an error in decoding the intended messages.
The sources $\{W_{i,t}\}$ can be regarded as messages by the source-channel
separation theorem. For example, to represent the broadcast channel
$p_{Y_{2},Y_{3}|X_{1}}$, let $W_{1}=(M_{1},M_{2})$, $Z_{2}=M_{1}$,
$Z_{3}=M_{2}$, where $M_{1},M_{2}$ are independent. Let $\mathcal{X}_{2}=\mathcal{X}_{3}=\mathcal{Y}_{1}=\mathcal{S}=\{0\}$
(i.e., $X_{2},X_{3},Y_{1},S$ are degenerate). Then the rate pair
$(R_{1},R_{2})$ is achievable for the broadcast channel if and only
if the joint source-channel Markov network admits a communication
scheme when $H(M_{1})=R_{1}$, $H(M_{2})=R_{2}$.\footnote{Technically, the capacity region is often defined as the closure of
the set of achievable $(R_{1},R_{2})$. We can represent the closure
operation in FOTPI by declaring the rate pair $(H(M_{1}),H(M_{2}))$
to be in the capacity region if for any $\tilde{M}_{1},\tilde{M}_{2}$
such that $H(\tilde{M}_{1})<H(M_{1})$, $H(\tilde{M}_{2})<H(M_{2})$
(which can be expressed using \eqref{rem:prev}), the network admits
a communication scheme for the messages $\tilde{M}_{1},\tilde{M}_{2}$.
For the purpose of simplicity, we ignore the closure operation.}

We will show that the capacity region of the joint source-channel
Markov network can be characterized in FOTPI. 

\smallskip{}

\begin{thm}
\label{thm:capacity}Fix $k\ge1$. There exists a first-order formula
\[
Q_{k}(W_{1},\ldots,W_{k},X_{1},\ldots,X_{k},Y_{1},\ldots,Y_{k},Z_{1},\ldots,Z_{k},S,L_{S},S',L_{S'})
\]
such that for any joint source-channel Markov network $(p_{W_{1},\ldots,W_{k}},p_{Y_{1},\ldots,Y_{k},S'|X_{1},\ldots,X_{k},S},\,p_{S},p_{Z_{1},\ldots,Z_{k}|W_{1},\ldots,W_{k}})$,
and any $p_{X_{1},\ldots,X_{k}}$ which assigns positive probability
for each $(x_{1},\ldots,x_{k})\in\mathcal{X}_{1}\times\cdots\times\mathcal{X}_{k}$,
the network admits a communication scheme if and only if the formula
$Q_{k}$ holds for 
\[
(X_{1},\ldots,X_{k},S,Y_{1},\ldots,Y_{k},S')\sim p_{X_{1},\ldots,X_{k}}p_{S}p_{Y_{1},\ldots,Y_{k},S'|X_{1},\ldots,X_{k},S}
\]
independent of 
\[
(W_{1},\ldots,W_{k},Z_{1},\ldots,Z_{k})\sim p_{W_{1},\ldots,W_{k}}p_{Z_{1},\ldots,Z_{k}|W_{1},\ldots,W_{k}},
\]
and $L_{S},L_{S'}$ are labels \eqref{eq:label_smid} of $S,S'$ respectively
(i.e., $\mathrm{label}_{3}(S,L_{S})$ and $\mathrm{label}_{3}(S',L_{S'})$
hold)\footnote{We require labels only for the present state $S$ and the next state
$S'$, but not for $W_{i},X_{i},Y_{i},Z_{i}$, since the channel $p_{Y_{1},\ldots,Y_{k},S'|X_{1},\ldots,X_{k},S}$
is essentially unchanged under relabelling of $W_{i},X_{i},Y_{i},Z_{i}$,
but not under relabelling of $S$ (while keeping $S'$ unmodified).
How the values of $S$ correspond to the values of $S'$ is essential.}.
\end{thm}
\begin{IEEEproof}
The main idea of the proof is to state the multi-letter operational
definition of the joint source-channel Markov network using a first-order
formula. Since the operational definition only consists of distribution
constraints, conditional independence constraints, and a bound on
the total variation distance, all of which can be expressed in first-order
formulae, the proof is a rather straightforward rewriting of these
constraints in the language of first-order formulae.

Let $W_{[k]}:=enc(W_{1},\ldots,W_{k})$ (with label $L_{W_{[k]}}$),
and define $X_{[k]},Y_{[k]},Z_{[k]}$ similarly. This can be checked
by
\begin{equation}
\mathrm{isseq}(W_{[k]},L_{W_{[k]}},k)\,\wedge\bigwedge_{i=1}^{k}\exists L.\,W_{i}\circ L\stackrel{as}{=}(W_{[k]}\circ L_{W_{[k]}})_{i},\label{eq:thm_defseq}
\end{equation}
and similar for $X_{[k]},Y_{[k]},Z_{[k]}$. Note that $k$ is fixed
and is not a variable, so the above formula is valid. 

We write $W_{i}^{n}=(W_{i,1},\ldots,W_{i,n})$. Let $W_{[k],i}:=enc(W_{1,t},\ldots,W_{k,t})$,
and $\bar{W}_{[k]}:=enc(W_{[k],1},\ldots,W_{[k],n})$ (with label
$\bar{L}_{\bar{W}_{[k]}}$). Define $\bar{X}_{[k]},\bar{Y}_{[k]},\bar{Z}_{[k]},\bar{\hat{Z}}_{[k]}$
similarly. Note that $\{Z_{i,t}\}_{i\in[k],t\in[n]}$ (used to define
$\bar{Z}_{[k]}$) are auxiliary random variables that do not appear
in the operational setting (they are not the same as $\hat{Z}_{i,t}$),
which will be used later in the bound on $P_{e}$. We check that they
are nested sequences by
\begin{align}
 & \mathrm{isseq}(\bar{W}_{[k]},\bar{L}_{\bar{W}_{[k]}},n)\,\nonumber \\
 & \wedge\,\forall t,\tilde{W},L_{\tilde{W}}.\big(\tilde{W}\circ L_{\tilde{W}}=(\bar{W}_{[k]}\circ)_{t}\,\to\,\mathrm{isseq}(W',L'_{W},k)\big),\label{eq:thm_wnest}
\end{align}
and similar for $\bar{X}_{[k]},\bar{Y}_{[k]},\bar{Z}_{[k]},\bar{\hat{Z}}_{[k]}$,
where we write $\bar{W}_{[k]}\circ=\bar{W}_{[k]}\circ\bar{L}_{\bar{W}_{[k]}}$
for brevity (when the label corresponding to the random variable is
clear from the context). We write $(\bar{W}_{[k]}\circ)_{t,i}:=((\bar{W}_{[k]}\circ)_{t})_{i}$
(which corresponds to $W_{i,t}$).

First we enforce that $(W_{1,t},\ldots,W_{k,t},Z_{1,t},\ldots,Z_{k,t})\sim p_{W_{1},\ldots,W_{k}}p_{Z_{1},\ldots,Z_{k}|W_{1},\ldots,W_{k}}$
i.i.d. across $t\in[n]$ by
\begin{align}
 & \forall t.\big(1\le t\le n\,\to(\bar{W}_{[k]}\circ)_{t}\stackrel{d}{=}W_{[k]}\circ\,\wedge\,(\bar{Z}_{[k]}\circ)_{t}|(\bar{W}_{[k]}\circ)_{t}\sim Z_{[k]}\circ|W_{[k]}\circ\nonumber \\
 & \;\;\;\;\;\;\wedge(\bar{W}_{[k]}\circ)_{t}(\bar{Z}_{[k]}\circ)_{t}\perp\!\!\!\perp(\bar{W}_{[k]}\circ)_{1..t-1}(\bar{Z}_{[k]}\circ)_{1..t-1}\big),\label{eq:thm_w}
\end{align}
where $(\bar{W}_{[k]}\circ)_{t}\stackrel{d}{=}W_{[k]}\circ$ means
\[
\exists W',L'.\big(W'\circ L'=(\bar{W}_{[k]}\circ\bar{L}_{\bar{W}_{[k]}})_{t}\,\wedge\,W'\circ L'\stackrel{d}{=}W_{[k]}\circ L_{W_{[k]}}\big),
\]
and refer to \eqref{eq:label_indep} for the meaning of $(\bar{W}_{[k]}\circ)_{t}(\bar{Z}_{[k]}\circ)_{t}\perp\!\!\!\perp(\bar{W}_{[k]}\circ)_{1..t-1}(\bar{Z}_{[k]}\circ)_{1..t-1}$.
Note that \eqref{eq:thm_w} checks that 
\[
(W_{1,t},\ldots,W_{k,t})\stackrel{d}{=}(W_{1},\ldots,W_{k}),
\]
\[
(Z_{1,t},\ldots,Z_{k,t})|(W_{1,t},\ldots,W_{k,t})\sim(Z_{1},\ldots,Z_{k})|(W_{1},\ldots,W_{k}),
\]
 and 
\[
(W_{1,t},\ldots,W_{k,t},Z_{1,t},\ldots,Z_{k,t})\perp\!\!\!\perp\{(W_{1,t'},\ldots,W_{k,t'},Z_{1,t'},\ldots,Z_{k,t'})\}_{t'<t}.
\]

Let $\bar{S}:=enc(S_{1},\ldots,S_{n+1})$ (with label $\bar{L}_{\bar{S}}$).
We check that $S_{1}\sim p_{S}$ is independent of $\{W_{i,t}\}_{i,t}$
by
\begin{align}
 & \mathrm{isseq}(\bar{S},\bar{L}_{\bar{S}},n+1)\,\wedge\,(\bar{S}\circ)_{1}\stackrel{d}{=}S\circ\,\wedge\,(\bar{S}\circ)_{1}\perp\!\!\!\perp\bar{W}_{[k]}.\label{eq:thm_s}
\end{align}
Write
\[
\tilde{X}\circ L_{\tilde{X}}=(\bar{X}_{[k]}\circ)_{1..t-1,i}
\]
for the $\tilde{X},L_{\tilde{X}}$ satisfying
\[
\mathrm{isseq}(\tilde{X},L_{\tilde{X}},t-1)\,\wedge\,\forall t'.\big(1\le t'\le t-1\,\to\,(\bar{X}_{[k]}\circ)_{t',i}=(\tilde{X}\circ)_{t'}\big),
\]
i.e., $\tilde{X},L_{\tilde{X}}$ corresponds to the sequence $X_{i,1},\ldots,X_{i,t-1}$.
At time $t\in[n]$, terminal $i$ outputs $X_{i,t}$ as a stochastic
mapping of $W_{i,1},\ldots,W_{i,n}$, $X_{i,1},\ldots,X_{i,t-1}$
and $Y_{i,1},\ldots,Y_{i,t-1}$ (conditionally independent of $\{W_{j,t'}\}_{j\in[k],t'\in[n]}$,
$\{X_{j,t'}\}_{j\in[k],t'<t}$, $\{Y_{j,t'}\}_{j\in[k],t'<t}$, $\{S_{j,t'}\}_{j\in[k],t'\le t}$
given those random variables) for $i\in[k]$. This is checked by
\begin{align}
 & \bigwedge_{i=1}^{k}\forall t.\big(1\le t\le n\,\to\nonumber \\
 & \;\;\;\;(\bar{X}_{[k]}\circ)_{t,i}\perp\!\!\!\perp(\bar{W}_{[k]}\circ)(\bar{X}_{[k]}\circ)_{1..t-1}(\bar{Y}_{[k]}\circ)_{1..t-1}(\bar{S}\circ)_{1..t}\nonumber \\
 & \;\;\;\;\;\;\big|\,(\bar{W}_{[k]}\circ)_{1..n,i}(\bar{X}_{[k]}\circ)_{1..t-1,i}(\bar{Y}_{[k]}\circ)_{1..t-1,i}\big).\label{eq:thm_x}
\end{align}
The channel generates $Y_{1,t},\ldots,Y_{k,t},S_{t+1}$ given $X_{1,t},\ldots,X_{k,t},S_{t}$
following $p_{Y_{1},\ldots,Y_{k},S'|X_{1},\ldots,X_{k},S}$. This
can be checked by
\begin{align}
 & \forall t.\big(1\le t\le n\,\to\nonumber \\
 & \;\;\;\;((\bar{Y}_{[k]}\circ)_{t},\,(\bar{S}\circ)_{t+1})\,|\,((\bar{X}_{[k]}\circ)_{t},\,(\bar{S}\circ)_{t})\sim(Y_{[k]}\circ,\,S'\circ)\,|\,(X_{[k]}\circ,\,S\circ)\nonumber \\
 & \;\;\;\;(\bar{Y}_{[k]}\circ)_{t}(\bar{S}\circ)_{t+1}\perp\!\!\!\perp(\bar{W}_{[k]}\circ)(\bar{X}_{[k]}\circ)_{1..t-1}(\bar{Y}_{[k]}\circ)_{1..t-1}(\bar{S}\circ)_{1..t-1}\nonumber \\
 & \;\;\;\;\;\;\big|\,(\bar{X}_{[k]}\circ)_{t}(\bar{S}\circ)_{t}\big),\label{eq:thm_y}
\end{align}
where the first line uses the notation $(X_{1}\circ L_{1},X_{2}\circ L_{2})$
in \eqref{eq:pair}. 

At the end, terminal $i$ outputs $\hat{Z}_{i,1},\ldots,\hat{Z}_{i,n}$
as a stochastic mapping of $W_{i,1},\ldots,W_{i,n}$, $X_{i,1},\ldots,X_{i,n}$
and $Y_{i,1},\ldots,Y_{i,n}$. This can be checked by
\begin{align}
 & \bigwedge_{i=1}^{k}\big((\bar{\hat{Z}}_{[k]}\circ)_{1..n,i}\perp\!\!\!\perp(\bar{W}_{[k]}\circ)(\bar{X}_{[k]}\circ)(\bar{Y}_{[k]}\circ)(\bar{S}\circ)\nonumber \\
 & \;\;\;\;\;\;\big|\,(\bar{W}_{[k]}\circ)_{1..n,i}(\bar{X}_{[k]}\circ)_{1..n,i}(\bar{Y}_{[k]}\circ)_{1..n,i}\big).\label{eq:thm_z}
\end{align}
By the coupling definition of total variation distance, we can enforce
$P_{e}\le\mathbf{P}(E)$ for some event $E$ using \eqref{eq:prle}
and the notation in \eqref{eq:event_notation} by
\begin{equation}
\mathrm{prle}\big(\{(\bar{Z}_{[k]}\circ)=(\bar{\hat{Z}}_{[k]}\circ)\},\,E\big).\label{eq:thm_pe}
\end{equation}
Finally, since the network admits a communication scheme if and only
if $P_{e}\le\mathbf{P}(E)$ is possible for any $\mathbf{P}(E)>0$,
the final formula $Q_{k}$ is
\begin{align*}
 & \forall E.\big(\mathrm{isev}(E)\,\wedge\,E\stackrel{\iota}{\neq}\emptyset\,\to\,\\
 & \;\;\exists n,W_{[k]},L_{W_{[k]}},X_{[k]},L_{X_{[k]}},Y_{[k]},L_{Y_{[k]}},Z_{[k]},L_{Z_{[k]}},\\
 & \;\;\;\;\bar{W}_{[k]},\bar{L}_{\bar{W}_{[k]}},\bar{X}_{[k]},\bar{L}_{\bar{X}_{[k]}},\bar{Y}_{[k]},\bar{L}_{\bar{Y}_{[k]}},\bar{Z}_{[k]},\bar{L}_{\bar{Z}_{[k]}},\bar{\hat{Z}}_{[k]},\bar{L}_{\bar{\hat{Z}}_{[k]}}.\\
 & \;\;(\cdots)\big),
\end{align*}
where the ``$\cdots$'' in the last line is the conjunction of \eqref{eq:thm_defseq}--\eqref{eq:thm_pe}.
\end{IEEEproof}
$ $

\medskip{}

\section{Definition of Single-Letter Characterization\label{sec:defslc}}

Theorem \ref{thm:capacity} shows that the capacity regions of a large
class of multiuser settings can be expressed as first-order formulae.
Arguably, this single-letter characterization is against the spirit
of single-letter characterizations in network information theory,
considering its complexity. Theorem \ref{thm:capacity} suggests that
general first-order formulae are perhaps too powerful, and focusing
on first-order ``single-letter'' formulae is not a meaningful restriction. 

As there was no generally accepted definition of single-letter characterization,
Körner \cite{korner1987concept} raised the question of finding a
logical theory on single-letter characterizations. After this, to
the best of the author's knowledge, the only attempt in providing
a definition of single-letter characterization is \cite{agarwal2018non},
which studies single-letter characterizations in the form of a conjunction
of linear inequalities on mutual information terms, where the alphabets
of all auxiliary random variables are fixed, which might be a little
restrictive (refer to Remark \ref{rem:prev}). In this section, we
propose some possible definitions that are (in a sense) more general
than \cite{agarwal2018non}, yet are more restrictive than general
first-order formulae. This may allow open problems regarding single-letter
characterizations of capacity regions to be stated in a rigorous manner.

We first define the probabilistic independence hierarchy in a similar
manner as the arithmetical hierarchy \cite{rogers1967theory} and
the Lévy hierarchy \cite{levy1965hierarchy}.
\begin{defn}
[The probabilistic independence hierarchy] A first-order formula
is in the set $\Delta_{0}^{\mathrm{pi}}=\Sigma_{0}^{\mathrm{pi}}=\Pi_{0}^{\mathrm{pi}}$
if and only if it is logically equivalent to a quantifier-free (i.e.,
all variables are free) formula. We define $\Delta_{i}^{\mathrm{pi}},\Sigma_{i}^{\mathrm{pi}},\Pi_{i}^{\mathrm{pi}}$
recursively. A first-order formula is in the set $\Sigma_{i+1}^{\mathrm{pi}}$
($i\ge0$) if and only if it is logically equivalent to a formula
in the form $\exists U_{1},\ldots,U_{k}.\,P(X_{1},\ldots,X_{n},U_{1},\ldots,U_{k})$
where $P$ is a formula in $\Pi_{i}^{\mathrm{pi}}$. A first-order
formula is in the set $\Pi_{i+1}^{\mathrm{pi}}$ ($i\ge0$) if and
only if it is logically equivalent to a formula in the form $\forall U_{1},\ldots,U_{k}.\,P(X_{1},\ldots,X_{n},U_{1},\ldots,U_{k})$
where $P$ is a formula in $\Sigma_{i}^{\mathrm{pi}}$. We then define
$\Delta_{i}^{\mathrm{pi}}=\Sigma_{i}^{\mathrm{pi}}\cap\Pi_{i}^{\mathrm{pi}}$.
\end{defn}
As a direct corollary of \cite{li2021undecidability}, $\Sigma_{3}^{\mathrm{pi}}$
and $\Pi_{3}^{\mathrm{pi}}$ are undecidable fragments of FOTPI. Finding
the lowest level in the probabilistic independence hierarchy which
is undecidable is left for future studies.
\begin{prop}
\label{prop:pi_frag}The problem of deciding the truth value of a
$\Sigma_{3}^{\mathrm{pi}}$ formula without free variables is undecidable.
The same is also true for $\Pi_{3}^{\mathrm{pi}}$.
\end{prop}
\begin{IEEEproof}
Note that $X\stackrel{\iota}{\le}Y$ \eqref{eq:lei} is $\Pi_{1}^{\mathrm{pi}}$,
$Z\stackrel{\iota}{=}XY$ \eqref{eq:joint} (which can be extended
to $Z\stackrel{\iota}{=}X_{1}\cdots X_{n}$) is $\Pi_{2}^{\mathrm{pi}}$,
$X\perp\!\!\!\perp Y|Z$ \eqref{eq:ci} is $\Sigma_{3}^{\mathrm{pi}}$,
and $\mathrm{card}_{=2}(X)$ \eqref{eq:cardeq} is $\Pi_{2}^{\mathrm{pi}}$.
The undecidability of $\Sigma_{3}^{\mathrm{pi}}$ follows from \cite[Cor. 3]{li2021undecidability},
which states that the problem of deciding whether there exists random
variables $X_{1},\ldots,X_{n}$ satisfying some conditional independence
constraints where $X_{1}$ is nondegenerate binary is undecidable.
The undecidability of $\Pi_{3}^{\mathrm{pi}}$ is due to the fact
that the negation of a $\Sigma_{3}^{\mathrm{pi}}$ formula is a $\Pi_{3}^{\mathrm{pi}}$
formula.
\end{IEEEproof}
\medskip{}

In FOTPI, the atomic formulae are probabilistic independence statements.
In network information theory, rate regions are more often stated
using linear inequalities on entropy terms. We can also define another
hierarchy with linear inequalities as atomic formulae.
\begin{defn}
[The linear entropy hierarchy] A first-order formula is in the set
$\Delta_{\mathrm{atom}}^{H}$ if and only if it is either in the form
$\mathbf{a}^{T}\mathbf{h}(X_{1},\ldots,X_{n})\ge0$ for some $\mathbf{a}\in\mathbb{Q}^{2^{n}-1}$
(where $\mathbf{h}(X_{1},\ldots,X_{n})\in\mathbb{R}^{2^{n}-1}$ is
the entropic vector \cite{zhang1997non} of $(X_{1},\ldots,X_{n})$),
or in the form $\tilde{Y}|\tilde{X}_{1},\ldots,\tilde{X}_{n}\stackrel{r}{\sim}Y|X_{1},\ldots,X_{n}$
(see Proposition \ref{prop:samecd}) \footnote{This is needed for channel coding problems to allow changing the input
distribution. For example, the capacity region of the point-to-point
channel $p_{Y|X}$ is $\exists\tilde{X},\tilde{Y}.\,\tilde{Y}|\tilde{X}\stackrel{r}{\sim}Y|X\,\wedge\,H(M)\le I(\tilde{X};\tilde{Y})$
(where the rate is $R=H(M)$). This is also useful for outer bounds
involving coupling, e.g. \cite{sato1978outer}.}. A first-order formula is in the set $\Delta_{0}^{H}=\Sigma_{0}^{H}=\Pi_{0}^{H}$
if and only if it is logically equivalent to a composition of $\Delta_{\mathrm{atom}}^{H}$
formulae using logical conjunction, disjunction and negation. We define
$\Delta_{i}^{H},\Sigma_{i}^{H},\Pi_{i}^{H}$ in a similar manner as
$\Delta_{i}^{\mathrm{pi}},\Sigma_{i}^{\mathrm{pi}},\Pi_{i}^{\mathrm{pi}}$. 
\end{defn}
Note that $\Sigma_{1}^{H}$ shares some similarities with \cite{agarwal2018non}
(refer to Remark \ref{rem:prev}). Since entropy and the $\stackrel{r}{\sim}$
relation can be defined in FOTPI, the linear entropy hierarchy is
in FOTPI as well. Similar to Proposition \ref{prop:pi_frag}, $\Sigma_{2}^{H}$
and $\Pi_{2}^{\mathrm{pi}}$ are undecidable fragments of FOTPI.
\begin{prop}
\label{prop:h_frag}The problem of deciding the truth value of a $\Sigma_{2}^{H}$
formula without free variables is undecidable. The same is also true
for $\Pi_{2}^{\mathrm{pi}}$. Also, the problem of deciding the truth
value of $P(X)$, where $P$ is a $\Sigma_{1}^{H}$ formula and $X\sim\mathrm{Bern}(1/2)$,
is undecidable. The same is also true for $\Pi_{1}^{\mathrm{pi}}$.
\end{prop}
\begin{IEEEproof}
Note that $\mathrm{card}_{=2}(X)$ \eqref{eq:cardeq} is $\Pi_{1}^{H}$.
The result follows from \cite[Cor. 3]{li2021undecidability}, which
states that the problem of deciding whether there exists $X_{1},\ldots,X_{n}$
satisfying some conditional independence constraints where $X_{1}$
is nondegenerate binary (also holds for the constraint $X_{1}\sim\mathrm{Bern}(1/2)$)
is undecidable. 
\end{IEEEproof}
\medskip{}

Theorem \ref{thm:capacity} states that for a general class of multiuser
coding settings, the capacity region can be stated as a first-order
formula, and hence is in the linear entropy hierarchy. 
\begin{prop}
The formula for the capacity region of the joint source-channel Markov
network in Theorem \ref{thm:capacity} is in $\Delta_{18}^{H}$.
\end{prop}
\begin{IEEEproof}
Note that $X\perp\!\!\!\perp Y$ can be expressed as $H(XY)-H(X)-H(Y)\ge0$,
so Theorem \ref{thm:capacity} is still valid when the atomic formulae
are linear inequalities on entropy terms. Technically, using linear
inequalities would restrict attention to random variables with finite
entropy, though it can be checked that the construction in Theorem
\ref{thm:capacity} does not require any random variable with infinite
entropy (as long as the input random variables have finite entropy).

By tracing the construction in the proof of Theorem \ref{thm:capacity}
and counting the depth of quantifier alternation, we find out that
the formula is in $\Pi_{17}^{H}\subseteq\Delta_{18}^{H}$.
\end{IEEEproof}
\medskip{}

Using the linear entropy hierarchy, we propose some possible definitions
of single-letter characterization, in decreasing order of generality.
\begin{enumerate}
\item \textbf{Any first-order formula in $\bigcup_{i\ge0}\Delta_{i}^{H}$.}
By Theorem \ref{thm:capacity}, the capacity regions of a large class
of multiuser settings can be expressed as first-order formulae. \medskip{}
\item \textbf{A first-order formula in $\Delta_{i}^{H}$, $\Sigma_{i}^{H}$
or $\Pi_{i}^{H}$ for a fixed $i\ge2$.} For example, we may restrict
attention to $\Pi_{2}^{H}$ to allow only existential formulae and
``for all, there exists'' formulae (e.g. \cite{wagner2008improved,gohari2010information,yu2018distortion,gohari2020outer}).
\medskip{}
\item \textbf{A first-order formula in $\Sigma_{1}^{H}$, i.e., an existential
formula.} The majority of capacity regions and bounds in network information
theory \cite{elgamal2011network} are $\Sigma_{1}^{H}$ formulae,
where all auxiliary random variables are existentially quantified.
However, restricting to $\Sigma_{1}^{H}$ does not give any provable
benefit on the ease of computation of the region, since Proposition
\ref{prop:h_frag} shows that such formula is undecidable even for
the simple input distribution $X\sim\mathrm{Bern}(1/2)$. Therefore,
restricting to $\Sigma_{1}^{H}$ formulae is merely a simplicity concern
rather than a computability concern.\medskip{}
\item \textbf{A first-order formula in $\Sigma_{1}^{H}$, together with
cardinality bounds for all existentially-quantified random variables.}
Cardinality bounds on auxiliary random variables are given for many
capacity regions in network information theory \cite{elgamal2011network},
which might allow more efficient computation. Technically, we should
restrict the cardinality bounds to be computable functions of the
cardinalities of the (non-auxiliary) random variables in the coding
setting. Considering that logarithm is involved in the definition
of entropy, whether the cardinality bounds can make the regions computable
depends on the solution to Tarski's exponential function problem \cite{tarski1951decision}
(also see \cite{gomez2014network,khamis2020decision}). \medskip{}
\item \textbf{A first-order formula in $\Delta_{0}^{H}=\Sigma_{0}^{H}=\Pi_{0}^{H}$.}
This would disallow auxiliary random variables, and hence is probably
too restrictive. Regions without auxiliary random variables such as
the Slepian-Wolf region \cite{slepianwolf1973a} are in $\Delta_{0}^{H}$.\medskip{}
\end{enumerate}
\begin{table}
\begin{centering}
\begin{tabular}{|c|c|}
\hline 
\multicolumn{2}{|c|}{$\Delta_{18}^{H}$ }\tabularnewline
\multicolumn{2}{|c|}{e.g. capacity region of any setting in Theorem \ref{thm:capacity}}\tabularnewline
\hline 
\multicolumn{2}{|c|}{$\vdots$}\tabularnewline
\hline 
\multicolumn{2}{|c|}{$\Delta_{3}^{H}$ }\tabularnewline
\hline 
$\Sigma_{2}^{H}$ & $\Pi_{2}^{H}$\tabularnewline
 & e.g. outer bounds in \cite{wagner2008improved,gohari2010information,yu2018distortion,gohari2020outer}\tabularnewline
\hline 
\multicolumn{2}{|c|}{$\Delta_{2}^{H}$}\tabularnewline
\hline 
$\Sigma_{1}^{H}$ & $\Pi_{1}^{H}$\tabularnewline
e.g. degraded broadcast channel \cite{bergmans1973random,gallager1974capacity}, & \tabularnewline
multiple access channel \cite{ahlswede1971multi,liao1972multiple,ahlswede1974capacity} & \tabularnewline
\hline 
\multicolumn{2}{|c|}{$\Delta_{1}^{H}$}\tabularnewline
\hline 
\multicolumn{2}{|c|}{$\Delta_{0}^{H}=\Sigma_{0}^{H}=\Pi_{0}^{H}$}\tabularnewline
\multicolumn{2}{|c|}{e.g. Slepian-Wolf \cite{slepianwolf1973a}}\tabularnewline
\hline 
\end{tabular}\medskip{}
\par\end{centering}
\caption{\label{tab:hierarchy}The linear entropy hierarchy and examples of
capacity regions and bounds at each level.}
\end{table}

Perhaps, instead of setting a fixed limit on which levels in the linear
entropy hierarchy counts as single-letter characterizations, we can
find the lowest level that can express the capacity region of a given
setting. Theorem \ref{thm:capacity} implies that such level always
exists. Given the linear entropy hierarchy, we can state open problems
in network information theory, e.g. for the broadcast channel, as
follows.

\medskip{}

\noindent \textbf{Open problem.} What is the lowest level in the linear
entropy hierarchy $\Delta_{i}^{H},\Sigma_{i}^{H},\Pi_{i}^{H}$ such
that the capacity region of the broadcast channel $p_{Y,Z|X}$ can
be expressed as a first-order formula (with free variables $X,Y,Z,W_{1},W_{2}$,
where $W_{i}$ represents the rate $R_{i}$ via $R_{i}=H(W_{i})$)
in that level?

\medskip{}

Theorem \ref{thm:capacity} implies that the capacity region is expressible
in $\Delta_{18}^{H}$ (and hence in $\Sigma_{18}^{H}$ and $\Pi_{18}^{H}$).
The remaining question is to find the lowest possible level. If the
conjecture that the $3$-auxiliary Marton's inner bound \cite{marton1979broadcast,gelfand1980capacity,liang2007broadcast}
is optimal is correct, then the lowest level would be $\Sigma_{1}^{H}$.
Some outer bounds for the broadcast channel are the UV outer bound
\cite{elgamal1979broadcast,nair2006outer} ($\Sigma_{1}^{H}$ formula,
suboptimal) and the J version of the UV outer bound \cite{gohari2020outer}
($\Pi_{2}^{H}$ formula, optimality unknown). 

We can also raise the same question for interference channel \cite{ahlswede1974capacity}.
The Han-Kobayashi inner bound \cite{han1981new} ($\Sigma_{1}^{H}$
formula) was shown to be suboptimal by Nair, Xia and Yazdanpanah \cite{nair2015sub}.
To the best of the author's knowledge, there is no candidate single-letter
characterization of the capacity region that is conjectured to be
optimal. Some outer bounds for the interference channel are given
in \cite{sato1977two} ($\Sigma_{1}^{H}$ formula), \cite{carleial1983outer}
($\Sigma_{1}^{H}$ formula), \cite{etkin2011analysis} ($\Sigma_{1}^{H}$
formula), and \cite{gohari2020outer} ($\Pi_{2}^{H}$ formula). Perhaps
the reason we are unable to give a $\Sigma_{1}^{H}$ candidate for
the capacity region is that the actual lowest possible level is $\Pi_{2}^{H}$
(or higher).

\medskip{}

\begin{rem}
\label{rem:prev} In \cite{agarwal2018non}, two slightly different
definitions of single-letter characterizations are given (specialized
to the case where there is only one rate $R$):
\begin{itemize}
\item \cite[eqn (1)]{agarwal2018non}: A formula which is the conjunction
of inequalities in the form $\beta R+\sum_{i}\alpha_{i}I(U_{A_{i}};U_{B_{i}}|U_{C_{i}})\le0$
(where $U_{A_{i}}=\{U_{a}\}_{a\in A_{i}}$, $A_{i},B_{i},C_{i}\subseteq[n]$,
$\beta,\alpha_{i}\in\mathbb{R}$ are computable; note that $\beta$
can be nonpositive) and polynomial constraints on the joint probability
mass function.
\item \cite[eqn (3)]{agarwal2018non}: A formula which is the conjunction
of inequalities in the form $R\le\sum_{i}\alpha_{i}I(U_{A_{i}};U_{B_{i}}|U_{C_{i}})$
and polynomial constraints on the joint probability mass function.
\end{itemize}
The proof of the non-existence of single-letter characterization for
the Markov channel in \cite{agarwal2018non} is performed on the second
definition, which is more restrictive than the first definition. While
\cite{agarwal2018non} shows that the optimal rate in the second definition
can be computed up to arbitrary precision, whether this also holds
for the first definition is not entirely clear (it may depend on the
solution to Tarski's exponential function problem \cite{tarski1951decision}). 

The first definition \cite[eqn (1)]{agarwal2018non} is closer to
$\Sigma_{1}^{H}$ in this paper. Still there are some important differences.
In \cite{agarwal2018non}, the alphabets of all random variables are
fixed, and general polynomial constraints on the joint probability
mass function are allowed. These are not the case for $\Sigma_{1}^{H}$.
The fixed alphabet assumption is crucial to the proof of the non-existence
of single-letter characterization in \cite{agarwal2018non} (since
fixing the alphabet transforms the problem into an existential problem
on real numbers, which are the entries of the joint probability mass
function). On the other hand, unlike \cite{agarwal2018non}, $\Sigma_{1}^{H}$
allows logical negation and disjunction, making the set of rate regions
expressible in $\Sigma_{1}^{H}$ closed under union. It is unclear
whether the Markov network in \cite{agarwal2018non}, or the joint
source-channel Markov network in this paper, has a capacity region
expressible in $\Sigma_{1}^{H}$.
\end{rem}
\medskip{}

\section{Extension to Continuous Random Variables\label{sec:continuous}}

In the previous sections, we assumed all random variables are discrete.
In this section, we consider an extension to general random variables
(in the standard probability space). It is unclear a priori whether
this extension will increase or decrease the interpretability strength.
For example, the true first-order theory of natural numbers is undecidable,
whereas the true first-order theory of real numbers is decidable \cite{tarski1951decision,seidenberg1954new}.
Hence, it is possible that a theory in a continuous setting is not
stronger than its discrete counterpart. Nevertheless, we will show
that the condition that a random variable is discrete can be defined
using a first-order formula, and hence the first-order theory of general
random variables is at least as expressive as that of discrete random
variables.

Previously, we assumed all random variables are defined on the same
standard probability space $([0,1],\mathcal{F},P)$. This is possible
since the standard probability space is rich enough to allow defining
any new discrete random variable in addition to an existing finite
collection of discrete random variables. This is no longer possible
for continuous random variables, since one random variable (e.g. the
random variable defined by the identity map) may already exhaust the
randomness in the space, and it is impossible to define any nondegenerate
random variable independent of it. Therefore, we have to assign slightly
different semantics to logical symbols, in order to allow extension
of the probability space when new random variables are introduced.

We define the first-order theory of general random variables. Assume
all random variables take value in $[0,1]$. The predicate $\psi(X_{1},\ldots,X_{n})$
should be interpreted as a predicate on the joint probability distribution
$P_{X_{1},\ldots,X_{n}}$ over $[0,1]^{n}$, i.e., $\psi(X_{1},\ldots,X_{n})$
is a shorthand for $\psi(P_{X_{1},\ldots,X_{n}})$. For a first-order
predicate $\psi(X_{1},\ldots,X_{n},Y)$ on $P_{X_{1},\ldots,X_{n},Y}$,
the formula $\exists Y.\psi(X_{1},\ldots,X_{n},Y)$ (which is a predicate
on $P_{X_{1},\ldots,X_{n}}$) means $\exists P_{X_{1},\ldots,X_{n},Y}.((P_{X_{1},\ldots,X_{n},Y})_{X_{1},\ldots,X_{n}}=P_{X_{1},\ldots,X_{n}}\,\wedge\,\psi(P_{X_{1},\ldots,X_{n},Y}))$,
that is, there exists joint distribution $P_{X_{1},\ldots,X_{n},Y}$
over $[0,1]^{n+1}$ such that its $(X_{1},\ldots,X_{n})$-marginal
$(P_{X_{1},\ldots,X_{n},Y})_{X_{1},\ldots,X_{n}}$ is $P_{X_{1},\ldots,X_{n}}$,
and $\psi(P_{X_{1},\ldots,X_{n},Y})$ holds. We define $\forall Y.\psi(X_{1},\ldots,X_{n},Y)$
similarly. For logical conjunction 
\[
\psi_{1}(X_{1},\ldots,X_{l},Y_{1},\ldots,Y_{m})\wedge\psi_{2}(X_{1},\ldots,X_{l},Z_{1},\ldots,Z_{n}),
\]
it is interpreted as 
\begin{align*}
 & \psi_{1}((P_{X_{1},\ldots,X_{l},Y_{1},\ldots,Y_{m},Z_{1},\ldots,Z_{n}})_{X_{1},\ldots,X_{l},Y_{1},\ldots,Y_{m}})\\
 & \wedge\psi_{2}((P_{X_{1},\ldots,X_{l},Y_{1},\ldots,Y_{m},Z_{1},\ldots,Z_{n}})_{X_{1},\ldots,X_{l},Z_{1},\ldots,Z_{n}}),
\end{align*}
which is a predicate on $P_{X_{1},\ldots,X_{l},Y_{1},\ldots,Y_{m},Z_{1},\ldots,Z_{n}}$.
Define logical disjunction similarly. We can therefore define any
first-order formula recursively (the base cases are $\exists Y.\psi(Y)$
and $\forall Y.\psi(Y)$ without any free variable, which are interpreted
as $\exists P_{Y}.\psi(P_{Y})$ and $\forall P_{Y}.\psi(P_{Y})$ respectively).

We now check that some of the formulae in the previous sections still
hold in the first-order theory of general random variables. The condition
$U\stackrel{\iota}{=}\emptyset$ (i.e., $U$ is almost surely a constant,
or $P_{U}$is a degenerate distribution) can be checked using the
same formula $U\stackrel{\iota}{=}\emptyset\,\Leftrightarrow\,U\perp\!\!\!\perp U$.
For general random variables, $X\stackrel{\iota}{\le}Y$ means there
exists a measurable function $f:[0,1]\to[0,1]$ such that $X=f(Y)$
with probability $1$. We can check that \eqref{eq:lei} 
\[
X\stackrel{\iota}{\le}Y\Leftrightarrow\,\forall U.\,\big(U\perp\!\!\!\perp Y\,\to\,U\perp\!\!\!\perp X\big)
\]
is still valid for general random variables. It is straightforward
to check that $X\stackrel{\iota}{\le}Y\Rightarrow\forall U.(U\perp\!\!\!\perp Y\to U\perp\!\!\!\perp X)$.
For the other direction, assume $X\stackrel{\iota}{\nleq}Y$. Then
there exists measurable $A\subseteq[0,1]$ such that $\mathbf{1}_{A}(X)$
is not almost surely a function of $Y$ (i.e., $\mathbf{P}(\mathbf{P}(X\in A|Y)\in\{0,1\})<1$).
Note that $0<\mathbf{P}(X\in A)<1$. Let 
\[
U|(X=x,Y=y)\sim\begin{cases}
\mathrm{Unif}[0,\mathbf{P}(X\in A|Y=y)] & \mathrm{if}\;x\in A\\
\mathrm{Unif}[\mathbf{P}(X\in A|Y=y),1] & \mathrm{if}\;x\notin A.
\end{cases}
\]
It is straightforward to check that $U\sim\mathrm{Unif}[0,1]$ is
independent of $Y$. Assume the contrary that $U$ is independent
of $X$, we have
\begin{align*}
\frac{1}{2}=\mathbf{E}[U|X\in A] & =\mathbf{E}\left[\frac{1}{2}\mathbf{P}(X\in A|Y)\,\Big|\,X\in A\right],
\end{align*}
which implies $\mathbf{P}(X\in A|Y)=1$ given $X\in A$ with probability
$1$. Also,
\begin{align*}
\frac{1}{2}=\mathbf{E}[U|X\notin A] & =\mathbf{E}\left[\frac{1}{2}+\frac{1}{2}\mathbf{P}(X\in A|Y)\,\Big|\,X\notin A\right],
\end{align*}
which implies $\mathbf{P}(X\in A|Y)=0$ given $X\notin A$ with probability
$1$. Therefore, $\mathbf{P}(\mathbf{P}(X\in A|Y)\in\{0,1\})=1$,
contradicting the assumption that $\mathbf{1}_{A}(X)$ is not almost
surely a function of $Y$.

It is straightforward to check that the formula \eqref{eq:cardlen}
for $\mathrm{card}_{\le n}(X)$ (which checks whether there exists
$S\subseteq[0,1]$ with $|S|\le n$ and $\mathbf{P}(X\in S)=1$) still
holds, and the formula \eqref{eq:label_smis} for single-mass indicator
still holds. Therefore, we can check whether the distribution of $X$
is atomless (i.e., $\mathbf{P}(X=x)=0$ for any $x$) by
\[
\mathrm{atomless}(X):=\lnot\exists U.\mathrm{smi}(X,U).
\]

Finally, we define the formula for discrete random variables. Since
any probability distribution can be decomposed into a mixture of a
discrete distribution and an atomless distribution, checking whether
$X$ is a discrete random variable is equivalent to checking if it
does not contain an atomless component, i.e., a measurable set $S\subseteq[0,1]$
such that $\mathbf{P}(X\in S)>0$ and $\mathbf{P}(X=x)=0$ for any
$x\in S$. This can be checked by the formula
\begin{align}
\mathrm{discrete}(X) & :=\lnot\exists V,W.\big(VW\stackrel{\iota}{\le}X\,\wedge\,\mathrm{card}_{=2}(V)\,\wedge\,\mathrm{card}_{=2}(W)\,\wedge\,\mathrm{card}_{=3}(VW)\nonumber \\
 & \;\;\;\;\;\wedge\,\lnot\exists U.(\mathrm{smi}(X,U)\,\wedge\,\mathrm{smi}(VWU,V)\,\wedge\,\mathrm{smi}(VWU,W))\big).\label{eq:discrete_formula}
\end{align}
To check this, we first show that if there exists $V,W$ satisfying
the above constraints, then $X$ has an atomless component. Since
$\mathrm{card}_{=3}(VW)$, we may assume $(V,W)\in\{(0,0),(1,0),(0,1)\}$.
Let $S$ be the set of values of $X$ corresponding to $(V,W)=(0,0)$.
Assume the contrary that $S$ is not an atomless component. Then there
exists $x\in S$ such that $\mathbf{P}(X=x)>0$. Let $U=\mathbf{1}\{X=x\}$.
It is straightforward to check that $U$ satisfies the constraints
in \eqref{eq:discrete_formula}, leading to a contradiction.

We then show that if $X$ has an atomless component, then there exists
$V,W$ satisfying \eqref{eq:discrete_formula}. Let $S\subseteq[0,1]$
such that $\mathbf{P}(X\in S)>0$ and $\mathbf{P}(X=x)=0$ for any
$x\in S$. Divide $S$ into three disjoint sets $S_{0},S_{1},S_{2}$
with positive probabilities. Let $S_{0,0}:=S_{0}$, $S_{0,1}:=S_{1}$,
$S_{1,0}:=[0,1]\backslash(S_{0,0}\cup S_{0,1})$. Let $(V,W)=(0,0)$
if $X\in S_{0,0}$, $(V,W)=(0,1)$ if $X\in S_{0,1}$, $(V,W)=(1,0)$
if $X\in S_{1,0}$. Assume the contrary that there exists $U$ satisfying
the constraints in \eqref{eq:discrete_formula}. Since $\mathrm{smi}(X,U)$,
we assume $U=\mathbf{1}_{\{x_{0}\}}(X)$, $\mathbf{P}(X=x_{0})>0$.
Since $\mathrm{smi}(VWU,V)$, $U$ is degenerate conditional on $(V,W)=(1,0)$,
and hence $x_{0}\notin S_{1,0}$ (note that $S_{1,0}\neq\{x_{0}\}$
since $S_{2}\subseteq S_{1,0}$ is atomless). Since $\mathrm{smi}(VWU,W)$,
$U$ is degenerate conditional on $(V,W)=(0,1)$, and hence $x_{0}\notin S_{0,1}$.
Hence we have $x_{0}\in S_{0,0}\subseteq S$, leading to a contradiction. 

As a result, the first-order theory of general random variables is
at least as expressive as the first-order theory of discrete random
variables (if we can impose $\mathrm{discrete}(X)$ on each variable
$X$ in the theory of general random variables, it reduces to the
theory of discrete random variables). Hence, the first-order theory
of general random variables is also algorithmically undecidable.

We remark that it is unclear whether Theorem \ref{thm:capacity} holds
for channels with continuous input or output alphabet. The above argument
only shows that the capacity of discrete channels can still be characterized
using a first-order formula with general random variables by the construction
in Theorem \ref{thm:capacity}.

\medskip{}

\section{Acknowledgement}

The author acknowledges support from the Direct Grant for Research,
The Chinese University of Hong Kong (Project ID: 4055133). The author
would like to thank Chandra Nair and Raymond W. Yeung for their invaluable
comments.

\medskip{}

\[
\]

\bibliographystyle{IEEEtran}
\bibliography{ref}

\end{document}